\documentclass[showpacs,aps,prb,twocolumn,superscriptaddress,preprintnumbers,amsmath,amssymb,floatfix,10pt]{revtex4-1}
\usepackage{graphicx}
\usepackage[tight]{subfigure}
\usepackage{dcolumn}
\usepackage{float}
\usepackage{bm}
\usepackage{enumerate}
\usepackage[english]{babel}
\usepackage{amsfonts}
\usepackage{amsthm}
\usepackage{amssymb}
\usepackage{siunitx}
\usepackage{calc}
\usepackage{color}
\usepackage{ifthen}
\usepackage{hyperref}
\usepackage{csquotes}
\usepackage{upgreek} 
\graphicspath{{}{figures/}}

\renewcommand{\v}[1]{\ensuremath{\boldsymbol{\mathbf{#1}}}} 
\newcommand{\uv}[1]{\ensuremath{\boldsymbol{\mathbf{\hat{#1}}}}} 
\newcommand{\e}[1]{\, \mathrm{e}^{#1}} 
\newcommand{\abs}[1]{\left| #1 \right|} 
\newcommand{\mean}[1]{\left\langle #1 \right\rangle} 
\newcommand{\grad}[1]{\v{\nabla} #1} 
\newcommand{\curl}[1]{\v{\nabla} \times #1} 
\newcommand{\cpi}{\uppi} 

\newcommand{\be}{\begin{equation}}
\newcommand{\ee}{\end{equation}}
\newcommand{\ba}{\begin{eqnarray}}
\newcommand{\ea}{\end{eqnarray}}
\newcommand{\beq}{\begin{equation}}
\newcommand{\eeq}{\end{equation}}
\newcommand{\bea}{\begin{eqnarray}}
\newcommand{\eea}{\end{eqnarray}}

\newcommand{\f}{\frac}

\renewcommand{\i}{\mathrm{i}} 
\newcommand{\inv}{^{-1}} 
\usepackage{inputenc}
\usepackage[draft,english]{fixme} 
\inputencoding{utf8}
\providecommand*{\diff}
	{\@ifnextchar^{\DIfF}{\DIfF^{}}}
\def\DIfF^#1{%
	\mathop{\mathrm{\mathstrut d}}%
		\nolimits^{#1}\gobblespace}
\def\gobblespace{%
	\futurelet\diffarg\opspace}
\def\opspace{%
	\let\DiffSpace\!%
	\ifx\diffarg(%
		\let\DiffSpace\relax
	\else
		\ifx\diffarg[%
			\let\DiffSpace\relax
		\else
			\ifx\diffarg\{%
				\let\DiffSpace\relax
			\fi\fi\fi\DiffSpace}
\providecommand*{\Diff}
	{\@ifnextchar^{\DIFF}{\DIFF^{}}}
\def\DIFF^#1{%
	\mathop{\mathcal{\mathstrut D}}%
		\nolimits^{#1}\gobblespace}			
				
\providecommand*{\groupSU}[1]{\ensuremath{\mathrm{SU}( #1 )}} 
\providecommand*{\groupU}[1]{\ensuremath{\mathrm{U}( #1 )}} 
\providecommand*{\groupO}[1]{\ensuremath{\mathrm{O}( #1 )}} 
\providecommand*{\pderiv}[3][]{\frac{\partial^{#1}#2}{\partial #3^{#1}}} 

\begin{document}
\title{Phase structure and phase transitions  in a three dimensional $\groupSU{2}$ superconductor} 

\author{Egil V. Herland}
\affiliation{Department of Physics, Norwegian University of Science and Technology, N-7491 Trondheim, Norway}

\author{Troels A. Bojesen}
\affiliation{Department of Physics, Norwegian University of Science and Technology, N-7491 Trondheim, Norway}

\author{Egor Babaev}
\affiliation{Physics Department, University of Massachusetts, Amherst, Massachusetts 01003, USA}
\affiliation{Department of Theoretical Physics, The Royal Institute of Technology, 10691 Stockholm, Sweden}

\author{Asle Sudb\o}
\affiliation{Department of Physics, Norwegian University of Science and Technology, N-7491 Trondheim, Norway}

\begin{abstract}
{We study the three dimensional \groupSU{2}-symmetric noncompact $\text{CP}^1$ model, with two charged matter fields 
coupled minimally to a noncompact Abelian gauge-field. The phase diagram and the nature of the phase transitions in 
this model have attracted much interest after it was proposed to describe an unusual continuous transition
associated with deconfinement of  spinons. Previously, it has been  demonstrated for various two-component gauge 
theories that  weakly first-order transitions  may appear as continuous ones of a new universality class in simulations 
of relatively large, but finite systems.  We have performed Monte-Carlo calculations on substantially larger 
systems sizes than those in previous works. We find   
that in some  area of the phase diagram where at finite sizes one gets signatures consistent with 
a single first-order transition, in fact there is a sequence of two phase transitions with an \groupO{3} paired phase 
sandwiched in between. We report (i) a  new estimate for the location of a bicritical point and (ii) the 
first resolution of bimodal distributions in energy histograms at relatively low coupling strengths. We perform a flowgram analysis of the
  direct transition line with rescaling of the linear system size in order to obtain a data collapse. The data collapses {up} to coupling constants where we find bimodal distributions in energy histograms.
}

\end{abstract}


\maketitle

\section{Introduction}

Recently, the $\text{CP}^1$ model consisting of two matter fields coupled to an Abelian gauge field
has been of great interest in condensed matter physics.
One of the sources of interest is the 
proposed  concept of deconfined quantum criticality (DQC).
It has been intensively debated as a possible novel paradigm for quantum 
phase transitions.\cite{Senthil_Science_2004, Senthil_PRB_2004, Sandvik_PRL_2007, Melko_PRL_2008, Lou_PRB_2009, Sandvik_PRL_2010, Banerjee_PRB_2010,
  Kaul_PRL_2012, Motrunich_PRB_2004, Kuklov_Ann_Phys_2006, Smiseth_PRB_2005, Kragset_PRL_2006, Motrunich_ArXiv_2008, Kuklov_PRL_2008, Jiang_J_Stat_Mech_2008, Isaev_J_Phys_Cond_Mat_2010, Nogueira_PRB_2007, Kaul_PRB_2008}
Such quantum criticality has been suggested to describe phase transitions that would not fit  into the Landau-Ginzburg-Wilson (LGW)
paradigm of a continuous (second-order) phase transition.\cite{Senthil_Science_2004, Senthil_PRB_2004} 
In particular, the continuous 
quantum phase transition from an antiferromagnetic N\'eel state into a paramagnetic valence-bond solid (VBS) state,\cite{Read_Sachdev_1989,Read_Sachdev_1990} 
does not agree with the LGW-description, according to which two phases with different broken symmetries generically are separated by a first-order phase 
transition. 
Recently, evidence for the DQC scenario has been claimed in studies of the so-called $J$-$Q$ model,\cite{Sandvik_PRL_2007} which is a Heisenberg model 
with additional higher-order spin interaction terms. Namely, it was suggested that high-precision Quantum Monte Carlo simulations of this model  support 
a continuous N\'eel -  VBS phase transition in accordance with the DQC scenario.
\cite{Sandvik_PRL_2007, Melko_PRL_2008, Lou_PRB_2009, Sandvik_PRL_2010, Banerjee_PRB_2010, Kaul_PRL_2012} 

It has been proposed that the critical field theory of a continuous N\'eel - VBS phase transition is the so-called noncompact $\text{CP}^1$ 
model ($\text{NCCP}^1$), with a \groupSU{2} symmetric field coupled to a noncompact \groupU{1} gauge field in three dimensions (3D).
\cite{Senthil_Science_2004, Senthil_PRB_2004, Motrunich_PRB_2004} Initial efforts on studying this effective model were focused on the 
special case where the \groupSU{2} symmetry was broken down to a \groupU{1}$\times$\groupU{1} symmetry, i.e., the easy-plane limit. For 
this case, a continuous phase transition was claimed.\cite{Motrunich_PRB_2004} However, in Ref.~\onlinecite{Kuklov_Ann_Phys_2006},  the existence 
of a paired phase in the \groupU{1}$\times$\groupU{1} easy-plane action was pointed out. (For earlier discussions of paired phases in various 
\groupU{1}$\times$\groupU{1} systems, see Refs.~\onlinecite{Babaev_Nucl_Phys_2004,Babaev_Nat_2004,Smorgrav_PRL_2005,Smiseth_PRB_2005}.) 
Furthermore, resorting to  mean-field theory arguments, it has been pointed out that at least in the vicinity of a paired state (in the 
parameter space of the model), the direct phase transition from a symmetric state to a state with broken \groupU{1}$\times$\groupU{1} symmetry, 
should be first-order.\cite{Kuklov_Ann_Phys_2006} Subsequent Monte-Carlo calculations have reported a weak first-order phase transition
for the easy-plane $\text{NCCP}^1$ model.\cite{Kuklov_Ann_Phys_2006, Kragset_PRL_2006} The so-called flowgram method has also been introduced  in Ref.~\onlinecite{Kuklov_Ann_Phys_2006},
specifically to characterize weak first-order phase transitions. 
Using this method, the direct phase transition from a symmetric state to a state with broken \groupU{1}$\times$\groupU{1} symmetry, 
has been claimed to be first-order for any non-zero value of the coupling constant. The phase transitions in the  easy-plane limit of the $\text{NCCP}^1$ model were also extensively studied in variety of other regimes 
in the context of two-component superconductors with independently conserved condensates.\cite{Babaev_PRB_2002, Babaev_Nucl_Phys_2004,Babaev_Nat_2004,Smiseth_PRL_2004, Smorgrav_PRL_2005,Smiseth_PRB_2005, Herland_PRB_2010}

For the \groupSU{2}-symmetric case, Monte Carlo computations have been performed in Ref.~\onlinecite{Motrunich_PRB_2004}. Here, a direct 
second-order phase transition was suggested, but the system sizes that were considered were quite small. In a subsequent paper,\cite{Motrunich_ArXiv_2008} 
an extensive study of the model was performed. In particular, for the direct transition line, a second-order phase transition was {claimed}.
At  higher couplings to the gauge field, it was suggested to turn into a first-order transition via a tricritical point. On the other hand, 
in  Ref.~\onlinecite{Kuklov_PRL_2008} (see also Ref.~\onlinecite{Kuklov_ArXiv_2008}), it was argued that the direct transition line 
is first-order. The  flowgram method employed in Ref.~\onlinecite{Kuklov_PRL_2008} showed no evidence for a tricritical point along the direct transition line. 
Rather, in this work the large-scale behavior at small couplings to the gauge field was found to be the same as for higher couplings, where  
indications of a first-order transition were seen by resolving a bimodal distribution in the energy histograms. Note that at the system sizes
studied in Ref.~\onlinecite{Kuklov_PRL_2008}, no bimodal distributions were resolved at small coupling constants. Nonetheless, in 
Ref.~\onlinecite{Kuklov_PRL_2008}, it was concluded that   even for weaker couplings, bimodal distributions indicative of a first-order
phase transition would emerge for large  enough system sizes. This conclusion was based on the similarity of scaling of various quantities
for large and small couplings, as evidenced by the flowgrams.

The weakness of the observed first-order phase transition, combined with the necessity of assessing the order of the phase transitions also in the limit of 
vanishingly small coupling strength, renders this problem computationally extremely demanding. In this work, we therefore examine the phase diagram of the 
$\text{NCCP}^1$ model at substantially larger systems sizes than what has been done in previous works.\cite{Motrunich_ArXiv_2008,Kuklov_PRL_2008} 
Performing the computations on larger systems allows us  to perform a very detailed investigation of the range of parameters where a paired phase is 
sandwiched in between the fully disordered and fully ordered state. This means that these two phases are separated, not by a direct transition,
but by two separate transitions.{ At small system sizes, these two separate phase transitions in fact give signatures
which would lead one to 
 conclude that the system features one single first-order phase transition. The existence of a 
paired phase sandwiched in between the fully ordered and disordered states emerges only when {one considers} large enough systems. Our study thus 
allows us to provide improved estimates for the location of a bicritical point where the direct transition line splits into two.

\section{Model}
The continuum $\text{NCCP}^1$ model is written as

\begin{gather}
  \label{eq:SU2_continuum_Z}
  Z = \int \Diff\v{\Psi}\Diff\v{\Psi}^{\dagger} \Diff\v{A} \ \e{-\beta H}, \\
  \label{eq:SU2_continuum_H}
  H = \frac{1}{2}\int \diff^3 \v{x} \left\{\abs{\left[\grad - \i e \v{A}(\v{x})\right] \v{\Psi}(\v{x})}^2 + \left[\curl \v{A}(\v{x})\right]^2\right\},
\end{gather}
where $\beta$ is the inverse temperature and $\v{\Psi}^{\dagger}(\v{x}) = (\psi_1^{*}(\v{x}), \psi_2^{*}(\v{x}))$ are two
complex fields that are coupled to a noncompact gauge field
$\v{A}(\v{x})$ with charge $e$. The fields $\psi_c(\v{x})$, $c \in \{
1,2\}$, obey the $\text{CP}^1$ constraint, $\abs{\v{\Psi}(\v{x})} = 1$.

The model can be mapped onto a nonlinear \groupO{3} $\sigma$ model
coupled to massive vector fields.\cite{Babaev_PRB_2002}
By introducing the fields,
\begin{gather}
  \label{change_I}
  \v{C}(\v{x})=\frac{\i}{2}\sum_c \left[\psi_c(\v{x})\nabla
    \psi_c^{*}(\v{x}) - \psi_c^{*}(\v{x})\nabla \psi_c(\v{x})\right] - e
  \v{A}(\v{x}),\\
  \v{n}(\v{x}) = \v{\Psi}^{\dagger}(\v{x}) \v{\sigma} \v{\Psi}(\v{x}),
  \label{change_II}
\end{gather}
where the components of $\v{\sigma}$ are the Pauli matrices, the $\text{NCCP}^1$  model (\ref{eq:SU2_continuum_Z}) can be rewritten as\cite{Babaev_PRB_2002}
\begin{align}
  H=&\f{1}{8}[ \partial_{\mu}\v{n}(\v{x})]^2+\f{1}{2} \left[\v{C}(\v{x})\right]^2 \nonumber\\
  &+\frac{1}{2e^2}\Bigg\{\epsilon_{\mu \nu \lambda}
  \Big[\partial_{\nu} C_{\lambda}(\v{x})\nonumber\\
  & - \f{1}{4}\v{n}(\v{x}) \cdot \partial_{\nu} \v{n}(\v{x})
  \times \partial_{\lambda} \v{n}(\v{x})\Big]\Bigg\}^2,
\label{SV}
\end{align}
where sum over repeated indices is assumed. The model represents an \groupO{3} nonlinear $\sigma$ model coupled to a massive vector 
field $\v{C}(\v{x})$. The latter represents a charged mode, and its mass is the inverse magnetic field penetration length. 
{At least for  sufficiently large values of electric charge coupling, the model 
can undergo a Higgs transition (where gauge field becomes massless)} without 
restoring simultaneously any  broken global symmetries. In that case, the remaining broken global symmetry is 
\groupO{3} {which is described by the order  parameter  $\v{n}(\v{x})$. }

If one introduces an easy-plane anisotropy for the vector field $\v{n}(\v{x})$, this would break the symmetry of the model 
to \groupU{1}$\times$\groupU{1}, and the separation of variables yields a neutral and a charged mode, the physics of which has been 
extensively studied.\cite{Babaev_PRB_2002, Babaev_Nucl_Phys_2004,Babaev_Nat_2004,Smiseth_PRL_2004, Smorgrav_PRL_2005,Smiseth_PRB_2005, Kuklov_Ann_Phys_2006,Herland_PRB_2010} 
However, there is one substantial difference in the case of \groupSU{2} symmetry. The charged and neutral sectors are coupled through 
the last term in Eq.~\eqref{SV}. Another difference compared to the \groupU{1}$\times$\groupU{1} case is that in two dimensions, stable 
singly quantized vortex lines do not exist in a type-II \groupSU{2} model (the same applies to vortex lines in three dimensions).\cite{Achucarro_Physics_Reports_2000} 
On the other hand, a type-I \groupSU{2} model has energetically stable counterparts of {ordinary} singly 
quantized type-I vortices. Since composite vortices are topological excitations which lead to the occurrence of paired states in 
\groupU{1}$\times$\groupU{1} systems, this aspect makes the phase diagram of \groupSU{2} theory an especially interesting problem 
to study.

In the Monte Carlo simulations, we employ a lattice realization of this model on a cubic lattice with size $L^3$ and with lattice constant 
$a = 1$. The fields $\psi_c(\v{x})$ are then defined on the vertices $\v{r} \in \{i\uv{x} + j \uv{y} + k \uv{z}|i,j,k \in \{1, \ldots, L\}\}$ 
of the lattice, $\psi_c(\v{x}) \rightarrow \psi_{c,\v{r}}$. For the first term in \eqref{eq:SU2_continuum_H}, we rescale the gauge field by 
$e\inv$ and invoke the gauge invariant lattice difference,
\begin{equation}
  \label{eq:gauge_invariant_lattice_difference}
  \left[\pderiv{}{x_{\mu}}- \i e A_{\mu}(\v{x})\right]\psi_c(\v{x})
  \rightarrow \psi_{c,\v{r}+\uv{\mu}}\e{-\i A_{\mu,\v{r}}} - \psi_{c,\v{r}},
\end{equation}
where $\mu \in \{x, y, z\}$ and $\v{r} + \uv{\mu}$ denotes the nearest-neighbor lattice point to vertex $\v{r}$ in the $\mu$-direction. The 
gauge field $A_{\mu,\v{r}}$ lives on the $(\v{r},\v{r}+\uv{\mu})$ links of the lattice. For the Maxwell term we get 
\begin{equation}
  \label{eq:lattice_Maxwell_action}
\left[\curl \v{A}(\v{x})\right]_{\mu} \to e\inv\sum_{\nu,\lambda} \epsilon_{\mu \nu \lambda}\Delta_\nu A_{\lambda, \v{r}},
\end{equation}
where $\Delta_\nu$ is the forward finite difference operator, $\Delta_\nu A_{\lambda, \v{r}} \equiv A_{\lambda, \v{r}+\uv{\nu}} - A_{\lambda, \v{r}}$, 
and $\epsilon_{\mu \nu \lambda}$ is the Levi-Civita symbol. In addition, by invoking the $\text{CP}^1$ constraint and discarding constant factors in 
the partition function $Z$, we obtain the following lattice realization of the $\text{NCCP}^1$ model:
\begin{gather}
  \label{eq:SU2_lattice_Z}
  Z = \int \Diff \v{A} \int_0^1 \Diff u \int_0^{2\cpi}
  \Diff\theta_1 \int_0^{2\cpi} \Diff\theta_2 \ \e{-\beta H}, \\
  \begin{aligned}
  \label{eq:SU2_lattice_H}
  H = \sum_{\v{r}, \mu}
  \Bigg[&-\sqrt{u_{\v{r}}}\sqrt{u_{\v{r}+\uv{\mu}}}\cos\left(\Delta_\mu \theta_{1,\v{r}}-A_{\mu,\v{r}} \right)\\ \nonumber
    &-\sqrt{1-u_{\v{r}}}\sqrt{1-u_{\v{r}+\uv{\mu}}}\cos\left(\Delta_\mu\theta_{2,\v{r}}-A_{\mu,\v{r}} \right)\\ \nonumber
    &+ \frac{1}{2e^2}\left(\sum_{\nu,\lambda} \epsilon_{\mu \nu \lambda}\Delta_\nu A_{\lambda, \v{r}}\right)^2\Bigg],
  \end{aligned}
\end{gather}
where $u_{\v{r}} = \abs{\psi_{1,\v{r}}}^2 = 1-\abs{\psi_{2,\v{r}}}^2$ and where $\abs{\psi_{c,\v{r}}}$ is the
amplitude and $\theta_{c,\v{r}}$ is the phase of the complex fields
$\psi_{c,\v{r}}$.

\section{Details of the Monte Carlo simulations}
The Monte Carlo simulations are performed on a cubic lattice with periodic boundary conditions in all 
directions and with size $L^3$ where $L \in \{8,\dots,96\}$. Up to $4.0 \cdot 10^7$ sweeps over the lattice 
were performed for the largest systems, while up to $1.0 \cdot 10^7$ sweeps were used for initial equilibration 
and initialization of the coupling distribution (see below).  
Monte-Carlo time-series were routinely inspected for equilibration. To test for ergodicity, typically $4$ 
independent large simulations were performed for the largest system sizes.
Histograms based on raw data and reweighted data were also compared for consistency. For most of the simulations, 
the parallel tempering (PT) algorithm was employed.\cite{Hukushima_JPSJ_1996, Earl_PCCP_2005,Katzgraber_ArXiv_2011} 
To be specific, we fix the coupling $e$ and perform the computations on a number of replicas (typically 
from 8 to 32 depending on the system size $L$ and the range of $\beta$ values) in parallel at different 
values of $\beta$. A Monte Carlo sweep consists of systematically traversing all lattice points with local 
trial moves of all six field variables by the Metropolis-Hastings algorithm.\cite{Metropolis_J_Chem_Phys_1953,
Hastings_Biometrika_1970} For $u_{\v{r}}$, the proposed new values are chosen with uniform probability within the interval 
$[0,1]$, and for $\theta_{c,\v{r}}$, the proposed new values are chosen with uniform probability within the interval 
$[0,2\cpi \rangle$. For the noncompact gauge field, the proposed new values are chosen within some limited 
increment (typically $[-\cpi/4,\cpi/4]$) from the old values.\footnote{In practice, we discretize the domain of the field variables 
into a large number of bins, $n_{\text{b}}$, in order to speed up the computations by the use of lookup tables. We use $n_{\text{b}}=501$ in the simulations, which we
believe to be sufficiently large to render the simulation results indistinguishable from the continuum $n_{\text{b}} \to \infty$ limit. Test simulations with other $n_\text{b}$ values support this claim.} There is no gauge fixing involved in the simulations. In 
addition to these local trial moves, the Monte Carlo sweep also includes a PT trial move of swapping replicas 
at neighboring $\beta$ values. 

All replicas were initially thermalized from an ordered or disordered start configuration. Then, initial
runs were performed in order to produce an optimal distribution of couplings for the simulation. In some 
cases, the set of couplings was found by measuring first-passage-times.\cite{Nadler_PRE_2008} In this
approach, the optimal set of couplings maximizes the flow of replicas in parameter space, essentially by shifting 
coupling values towards the bottlenecks.\cite{Katzgraber_J_Stat_Mech_2006} However, in cases with no severe bottleneck, 
the optimal set of couplings was found by demanding that the acceptance rates for swapping neighboring replicas were 
equal for all couplings.\cite{Hukushima_PRE_1999} Irrespective of how the set of couplings was found, it was always 
ascertained that replicas were able to traverse parameter space sufficiently many times during production runs. The 
measurements were postprocessed by multiple histogram reweighting.\cite{Ferrenberg_PRL_1989} Random numbers were 
generated by the Mersenne-Twister algorithm.\cite{Matsumoto_1998_ACM_TMCS} Errors were determined by the jackknife 
method.\cite{Berg_Comput_Phys_Commun_1992}

As mentioned in the introduction, the $\text{NCCP}^1$ model is a difficult model on which to perform Monte Carlo computations.
In Ref.~\onlinecite{Kuklov_PRL_2008}, the $\text{NCCP}^1$ model was mapped to a so-called $J$-current model, which allows 
simulations based on the worm algorithm.\cite{Prokofev_PLA_1998,Prokofev_PRL_2001} (For the sake of completeness, and since to 
our knowledge the details of the mapping have not been published, we present the derivation of this mapping in
Appendix \ref{app:J_current_formulation}.) An approach based on the $J$-current model was attempted as well.
However, due to the presence of long-range interactions in this formulation, it was difficult to work with the lattice sizes above $L \sim 40$.
 Hence, the computations were performed on the model in the original $\text{NCCP}^1$ formulation, using the PT
algorithm with which it is easy to grid-parallelize the lattice.

\section{Observables and finite-size scaling}
Perhaps the most familiar quantity that is used to explore phase transitions, is the specific heat $C_v$. The specific 
heat is given by the second moment of the action, 
\begin{equation}
  \label{eq:specific_heat}
  C_v = \frac{\beta^2}{L^3}\mean{\left(H-\mean{H}\right)^2},
\end{equation}
where brackets $\langle \dots \rangle$ denote statistical averages. In most cases, $C_v$ exhibits a well-defined peak at the
phase transition. For a continuous phase transition the correlation length diverges with critical exponent $\nu$ as $\xi \sim
\abs{t}^{-\nu}$, with $t = (\beta-\beta_{\text{c}})/\beta$ being the deviation from the critical coupling $\beta_{\text{c}}$. The 
critical exponent $\alpha$ is defined by the singular part of $C_v$, given by $C_v \sim \abs{t}^{-\alpha}$. Then, in a limited 
system of size $L^3$, the finite-size scaling (FSS) of the specific heat is given by
\begin{equation}
  \label{eq:specific_heat_fss_continuous}
  C_v \sim C_0 + C_1L^{\alpha/\nu},
\end{equation}
where $C_0$ and $C_1$ are non-universal coefficients. For a first-order transition, with two coexisting phases and no diverging
correlation length, there is {indeed no  critical behavior}. Still, first-order transitions exhibit
well-behaved FSS with ``effective" exponents, $\alpha = 1$ and $\nu = 1/3$.\cite{Fisher_PRB_1982, Cardy_PRB_1983} Hence, the peak 
of the specific heat scales as 
\begin{equation}
  \label{eq:specific_heat_fss_first_order}
  C_v \sim L^{3},
\end{equation}
for a first-order transition. Distinguishing between continuous and first-order transitions is an important issue in the present 
work. For that purpose, FSS of the specific heat peak will play an important role. 

We also investigate the third moment of the action given by\cite{Sudbo_PRL_2002, Smiseth_PRB_2003}
\begin{equation}
  \label{eq:third_moment}
  M_3 = \frac{\beta^3}{L^3}\mean{\left(H-\mean{H}\right)^3}.
\end{equation}
In the vicinity of the critical point, this quantity typically features a minimum point and a maximum point [see for instance the 
inset in panel (b) of Fig.~\ref{fig:O3_scaling}]. The difference in the $M_3$ value of these two extrema scales as 
\begin{equation}
  \label{eq:third_moment_height_fss_continuous}
  (\Delta M_3)_{\text{height}} \sim L^{(1+\alpha)/\nu},
\end{equation}
and the difference in the coupling values scales as
\begin{equation}
  \label{eq:third_moment_width_fss_continuous}
  (\Delta M_3)_{\text{width}} \sim L^{-1/\nu},
\end{equation}
for a continuous phase transition. For a first-order transition, the FSS is
\begin{equation}
  \label{eq:third_moment_height_fss_first_order}
  (\Delta M_3)_{\text{height}} \sim L^{6},
\end{equation}
and
\begin{equation}
  \label{eq:third_moment_width_fss_first_order}
  (\Delta M_3)_{\text{width}} \sim L^{-3}.
\end{equation}

As was mentioned above, one may construct a three component gauge neutral field $\v{n}_{\v{r}}$,\cite{Babaev_PRB_2002} given by
\begin{equation}
  \label{eq:O3_vector}
  \v{n}_{\v{r}} = \v{\Psi}_{\v{r}}^{*} \v{\sigma} \v{\Psi}_{\v{r}},
\end{equation}
where the components of $\v{\sigma}$ are the Pauli matrices. Since it is a unit \groupO{3} vector, we can introduce a ``magnetization",
\begin{equation}
  \label{eq:O3_magnetization}
  \v{M} = \sum_{\v{r}} \v{n}_{\v{r}}.
\end{equation}
The order parameter $\mean{\v{m}}$, where $\v{m} = \v{M}/L^3$, signals the
onset of order in the $\groupO{3}$ gauge neutral vector field
$\v{n}_{\v{r}}$, and the critical point of this transition can be accurately
determined by a proper analysis of the finite-size crossings of the
associated Binder cumulant,\cite{Binder_PRL_1981, Binder_Z_Phys_B_1981,
Sandvik_AIP_2010}
\begin{equation}
  \label{eq:O3_Binder_cumulant}
  U_4 = \frac{5}{2} - \frac{3\langle M^4 \rangle}{2\langle M^2 \rangle^2}.
\end{equation}
The finite-size crossings of the Binder cumulant are known to converge rapidly towards
the critical coupling $\beta_{\text{c}}$. Hence, $\beta_{\text{c}}$ can be accurately 
determined by a simple extrapolation of the finite-size crossings to the thermodynamic 
limit or by invoking scaling forms that account for finite-size corrections.
\cite{Beach_ArXiv_2005, Sandvik_AIP_2010}  

A number of quantities related to magnetization may be used to extract critical exponents from the 
Monte Carlo simulations. The magnetic susceptibility, given by
\begin{equation}
  \label{eq:O3_susceptibility}
  \chi = L^3 \beta \mean{m^2},
\end{equation}
when $\beta < \beta_{\text{c}}$, scales as $\chi \sim L^{2-\eta}$ at $\beta =
\beta_{\text{c}}$. Hence, we may determine the anomalous scaling dimension
$\eta$ by FSS of $\chi$ measurements obtained at $\beta_{\text{c}}$.

The exponent $\nu$ can, alternatively, be determined by calculating the logarithmic derivative of the second power of the 
magnetization,\cite{Ferrenberg_PRB_1991} 
\begin{equation}
  \label{eq:O3_logarithmic_derivative}
  \pderiv{}{\beta}\ln\mean{m^2} = \frac{\mean{m^2 H}}{\mean{m^2}} - \mean{H}.
\end{equation}
The FSS of this quantity is $\pderiv{}{\beta}\ln\mean{m^2} \sim L^{1/\nu}$. Since the logarithmic derivative 
exhibits a peak that is associated with the critical point, it is possible to extract $\nu$ by measuring the 
logarithmic derivative at the pseudocritical point, without an accurate determination of $\beta_{\text{c}}$.

Similar to Ref.~\onlinecite{Motrunich_ArXiv_2008}, we {search for} the critical point of the Higgs transition by measuring the dual stiffness
\begin{equation}
  \label{eq:Higgs_dual_stiffness}
  \rho_{\text{dual}}^{\mu \mu}(\v{q}) = 
  \mean{\frac{\abs{\sum_{\v{r},\nu,\lambda}\epsilon_{\mu \nu \lambda}\Delta_\nu A_{\lambda, \v{r}}\e{\i\v{q}\v{r}}}^2}{(2\cpi)^2L^3}},
\end{equation}
which is the Fourier space correlator of the magnetic field. This order parameter for the Higgs transition 
is dual in the sense that it is finite in the high-temperature phase and zero in the low-temperature phase. Like
 in Ref.~\onlinecite{Motrunich_ArXiv_2008}, this quantity is measured at the smallest available wavevector 
$\v{q} \neq \v{0}$. We chose to measure $\rho_{\text{dual}}^{zz}$ at $\v{q}_{\text{min}} = (2\cpi/L,0,0)$. At the 
critical point, the quantity $L\rho_{\text{dual}}^{\mu \mu}(\v{q}_{\text{min}})$ is universal, such that the finite-size 
crossings of $L\rho_{\text{dual}}^{\mu \mu}(\v{q}_{\text{min}})$ can be used to estimate the critical point of the Higgs 
transition. In addition, measuring the coupling derivative of $L\rho_{\text{dual}}^{\mu \mu}(\v{q}_{\text{min}})$ can be used 
to estimate the correlation length exponent $\nu$ as 
\begin{equation}
  \label{eq:Lrho_coupling_derivative_scaling}
  \pderiv{}{\beta}L\rho_{\text{dual}}^{\mu \mu}(\v{q}_{\text{min}}) \sim L^{1/\nu},
\end{equation}
at the critical point.

\section{Numerical results}

\subsection{Outline of the phase diagram}

The phase diagram of the $\text{NCCP}^1$ model is presented in Fig.~\ref{fig:Phase_diagram}. For small values of 
$\beta$, there is a normal phase that can be recognized by a disordered gauge neutral vector field $\v{n}_{\v{r}}$
and a massless gauge field. Hence, $\mean{m} = 0$ and $\rho_{\text{dual}}^{\mu \mu}(\v{q}) \neq 0$ in this phase. 
For large values of $e$ and higher values of $\beta$, there is a transition into a phase that we label the \groupO{3} 
phase. Here, the vector field $\v{n}_{\v{r}}$ is ordered (the \groupO{3} symmetry is spontaneously broken),
$\mean{m} \neq 0$, whereas the gauge field remains massless, $\rho_{\text{dual}}^{\mu \mu}(\v{q}) \neq 0$. In the 
case of $\groupU{1} \times \groupU{1}$ symmetric superconductors, this phase is sometimes denoted a 
\textit{metallic superfluid} or a \textit{paired phase}, with long-range 
order in the gauge neutral linear combination of the phases (in the $\groupU{1} \times \groupU{1}$ case), but not in the individual
ones.\cite{Babaev_Nucl_Phys_2004, Babaev_Nat_2004, Babaev_PRL_2005, Smiseth_PRB_2005,
  Smorgrav_PRL_2005, Kuklov_Ann_Phys_2006, Herland_PRB_2010} 
 From the
\groupO{3} phase, by reducing the value of $e$, one enters an ordered phase that we label
the \groupSU{2} phase. Going into this phase, the gauge field
dynamically acquires a Higgs mass and the system becomes a
two-component $\text{NCCP}^1$  superconductor. Note that the Higgs transition is
related to a local  symmetry, and indeed, is not associated with spontaneous symmetry
breaking.\cite{Elitzur_PRD_1975}  This aspect should be kept in mind where we for brevity refer
to the fully ordered state as ``broken \groupSU{2}'' or ``fully broken state'' to distinguish it 
from a paired state. The \groupSU{2} phase is recognized by measuring $\mean{m} \neq 0$ 
and $\rho_{\text{dual}}^{\mu  \mu}(\v{q}) = 0$.  

It is generally expected that at small values of $e$, 
the \groupSU{2} phase may also be entered directly from the normal phase, i.e., without going through the 
intermediate paired phase. The nature of the phase transition along this direct transition line {in this and related multicomponent models has been}  intensively 
debated due to its relevance to deconfined quantum criticality. We will return to the direct transition line in 
Secs.~\ref{Subsec:Bicritical_region} and \ref{Subsec:flowgram}. First, we present results for the two 
separate transition lines. 

\begin{figure}[tbp]
\includegraphics[width=\columnwidth]{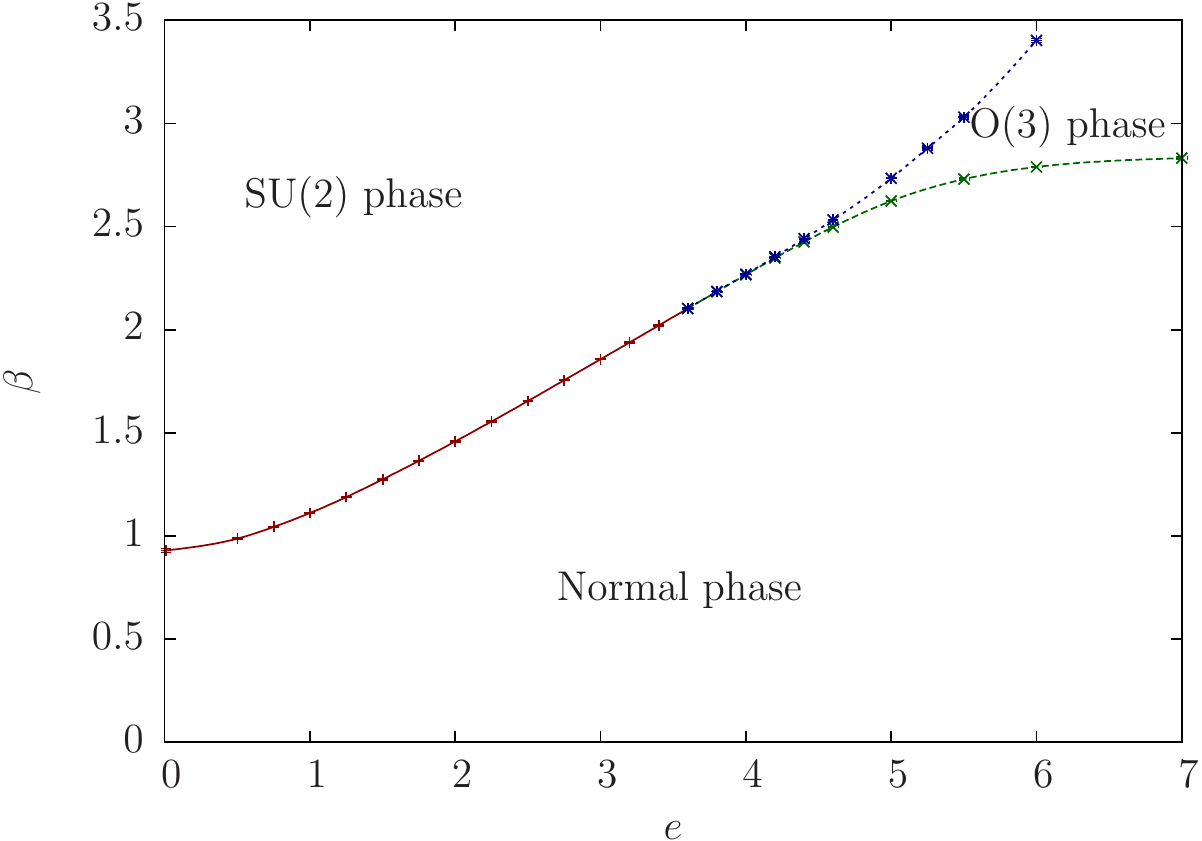}
\caption{(Color online) Phase diagram of the $\text{NCCP}^1$
  model. \groupSU{2} phase: Fully ordered phase where the \groupO{3}
  symmetry is spontaneously broken, $\mean{m} \neq 0$, and the gauge
  field is massive, $\rho_{\text{dual}}^{\mu \mu}(\v{q}) =
  0$. \groupO{3} phase: \groupO{3} symmetry is spontaneously broken,
  $\mean{m} \neq 0$, but the gauge field is massless,
  $\rho_{\text{dual}}^{\mu \mu}(\v{q}) \neq 0$. Normal phase: \groupO{3} symmetry is restored,
  $\mean{m} = 0$, and the gauge field is massless,
  $\rho_{\text{dual}}^{\mu \mu}(\v{q}) \neq 0$. The direct transition
  line from the \groupSU{2} phase to the normal phase is denoted by
  $+$-markers and a solid red line. The Higgs transition line
  between the \groupSU{2} phase and the \groupO{3} phase is denoted by
  $\ast$-markers and a dotted blue line. The transition line
  between the \groupO{3} phase and the normal phase is denoted by
  $\times$-markers and a dashed green line. Lines are guide to the eyes.
}
\label{fig:Phase_diagram}
\end{figure}

\subsubsection{O(3) line}
In Refs.~\onlinecite{Kuklov_PRL_2008,Kuklov_nordita,Kuklov_APS_2008}, the existence of an intermediate paired 
phase, separating a fully ordered state from a fully disordered one, was shown in the \groupSU{2}-symmetric 
theory. The nonlinear $\sigma$ model mapping presented above, suggests that the transition 
line between the normal phase and the \groupO{3} phase should  be a continuous transition in the \groupO{3} 
universality class, at least in the limit far from the bicritical point. 
We have considered this for the case $e = 6.0$, and the FSS results are given in Fig.~
\ref{fig:O3_scaling}. A log-log plot of the FSS of the peak height in $\pderiv{}{\beta}\ln\mean{m^2}$ is 
given in panel (a), and the measured peak heights fall on a straight
line for $L \geq 20$. The best fit to the form $\pderiv{}{\beta}\ln\mean{m^2} \sim
L^{1/\nu}$ yields $\nu = \num{0.715 \pm 0.004}$. In panel (b), we also measure $(\Delta
M_3)_{\text{height}}$, and this quantity exhibits negligible finite-size corrections to scaling at least 
for $L \geq 10$. The best fit according to Eq.~\eqref{eq:third_moment_height_fss_continuous} yields $\alpha =
\num[separate-uncertainty=true]{-0.117 \pm 0.011}$, where the value of $\nu$ obtained above was used. In this case, 
it was found that $\nu$ was most precisely determined by measuring the peak height in $\pderiv{}{\beta}\ln\mean{m^2}$ 
rather than measuring $(\Delta M_3)_{\text{width}}$. The maximum peak in $M_3$ is not very sharp [see the inset of 
panel (b)]. Thus, the error bars in $(\Delta M_3)_{\text{width}}$ are large. In order to determine $\eta$, the FSS of 
the magnetic susceptibility $\chi$ is given in panel (c). Here, $\chi$ is measured at the critical coupling
$\beta_{\text{c}} = \num{2.7894 \pm 0.0003}$, which was determined by fitting the Binder crossings 
of $L$ and $L/2$ to a function that accounts for power-law finite-size corrections. The best fit of 
$\chi(L)$ was determined for sizes $L \in \{12, \dots, 64\}$ to yield $\eta = \num{0.024 \pm 0.014}$. All 
the exponents listed above correspond well with the exponents of the \groupO{3} universality 
class.\cite{Holm_PRB_1993, Campostrini_PRB_2002}

\begin{figure}[tbp]
\includegraphics[width=0.9\columnwidth]{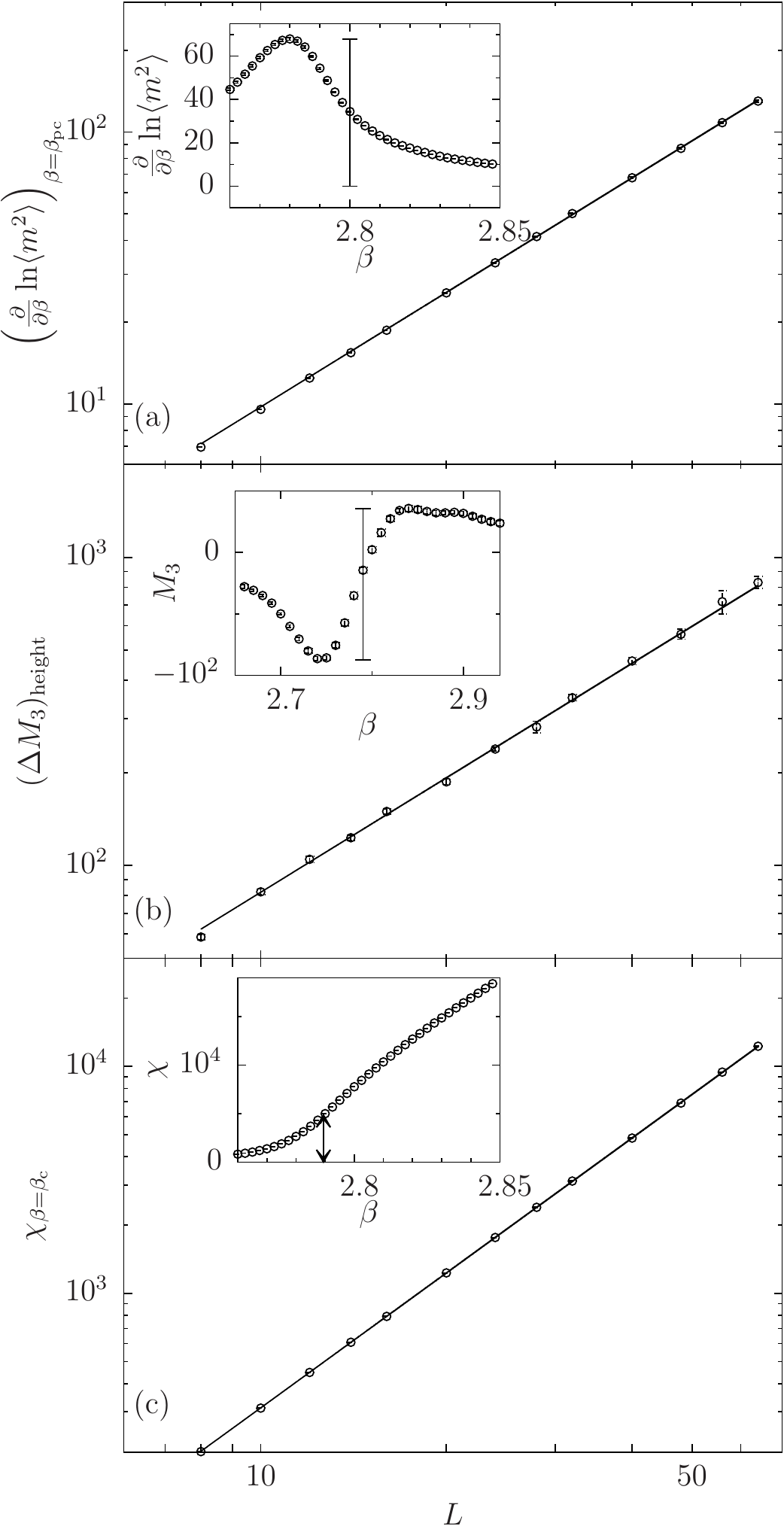}
\caption{FSS results for the transition between
  the normal phase and the \groupO{3} phase when $e = 6.0$. 13 system
  sizes $L \in \{8, \dots, 64\}$ are used. In all panels, the
  solid straight line is the best fit obtained for a fitting function on the
  form $aL^b$ with two free parameters $a$ and $b$. Panel (a): Log-log
  plot of the maximum in the logarithmic derivative of the second
  power of the magnetization
  $(\partial/\partial\beta \ln\langle m^2 \rangle)_{\beta = \beta_{\text{pc}}}$ [see
  Eq.~\eqref{eq:O3_logarithmic_derivative}] as a function of $L$. The
  best fit is obtained for sizes $L \in \{20, \dots, 64\}$. The inset shows
  the measure of $(\partial/\partial\beta \ln\langle m^2
  \rangle)_{\beta = \beta_{\text{pc}}}$ in the case when $L = 40$. Panel (b): Log-log
  plot of the third moment height difference $(\Delta
  M_3)_{\text{height}}$ as a function of $L$. The best fit is obtained
  for sizes $L \in \{10, \dots, 64\}$. The inset shows the measure $(\Delta M_3)_{\text{height} }$ in
  the case when $L = 14$. Panel (c): Log-log
  plot of the magnetic susceptibility measured at the critical
  coupling $\chi_{\beta = \beta_{\text{c}}}$ as a function of $L$. The
  best fit is obtained for sizes $L \in \{12, \dots, 64\}$. The inset shows
  $\chi_{\beta = \beta_{\text{c}}}$ for the case when $L =
  40$, and the arrowheads indicate that $\chi$ is measured at the same fixed
  coupling $\beta_{\text{c}}$ for all sizes.}
\label{fig:O3_scaling}
\end{figure}

\subsubsection{Superconducting transition.}
Computations have also been performed along the transition line between the \groupO{3} phase and the \groupSU{2} phase. In analogy with the 
paired phase of the \groupU{1}$\times$\groupU{1} model
\cite{Babaev_Nucl_Phys_2004,Babaev_Nat_2004,Smiseth_PRL_2004, Smorgrav_PRL_2005,Smiseth_PRB_2005, Kuklov_Ann_Phys_2006,Herland_PRB_2010}  
(i.e., the metallic superfluid), the transition to the \groupO{3} sector should be associated with the proliferation of  single-quanta 
vortices. In the \groupU{1}$\times$\groupU{1} model, such vortices have similar phase windings in both complex fields and are topologically 
well-defined objects. In the \groupSU{2} case, such vortices can have either similar phase windings in both components, or a phase winding 
only in one component if the other component exists only in the vortex core of the former. Such objects are non-topological, and are 
unstable in type-II \groupSU{2} superconductors.\cite{Achucarro_Physics_Reports_2000} This suggests that the system should be a type-I \groupSU{2} 
superconductor in order to feature a phase transition into a paired phase. In analogy with single-component
type-I superconductors, one would then expect a first-order phase transition.\cite{Halperin_Lubensky_Ma_1974,Mo_Hove_Sudbo_2001} A different viewpoint is based on mean-field arguments, which suggest that the transition line could be a first-order transition
line in the vicinity of a bicritical point.\cite{Kuklov_Ann_Phys_2006} Other objects which can disorder the Higgs sector, are Hopfions\cite{Babaev_PRB_2002,Babaev_PRB_2009}. In this work we have made no serious attempts at resolving such topological defects.

To check the universality class of this line, FSS results of $\partial / \partial \beta [L\rho_{\text{dual}}^{z z}(\v{q}_{\text{min}})]$, obtained 
at the critical point with $e = 5.0$, are given in Fig.~\ref{fig:Inv_U1_scaling}. First, the critical coupling was determined to be
$\beta_{\text{c}} = \num{2.7347 \pm 0.0005}$, by considering the crossings of $L\rho_{\text{dual}}^{zz}(\v{q}_{\text{min}})$ (see the
inset of Fig.~\ref{fig:Inv_U1_scaling}). Then, the correlation length exponent was estimated to be $\nu = \num{0.664 \pm 0.039}$. 
This value is consistent with an inverted 3Dxy transition line.\cite{Campostrini_PRB_2001}  
We have not been able to resolve a first-order phase transition at this line.

\begin{figure}[tbp]
\includegraphics[width=\columnwidth]{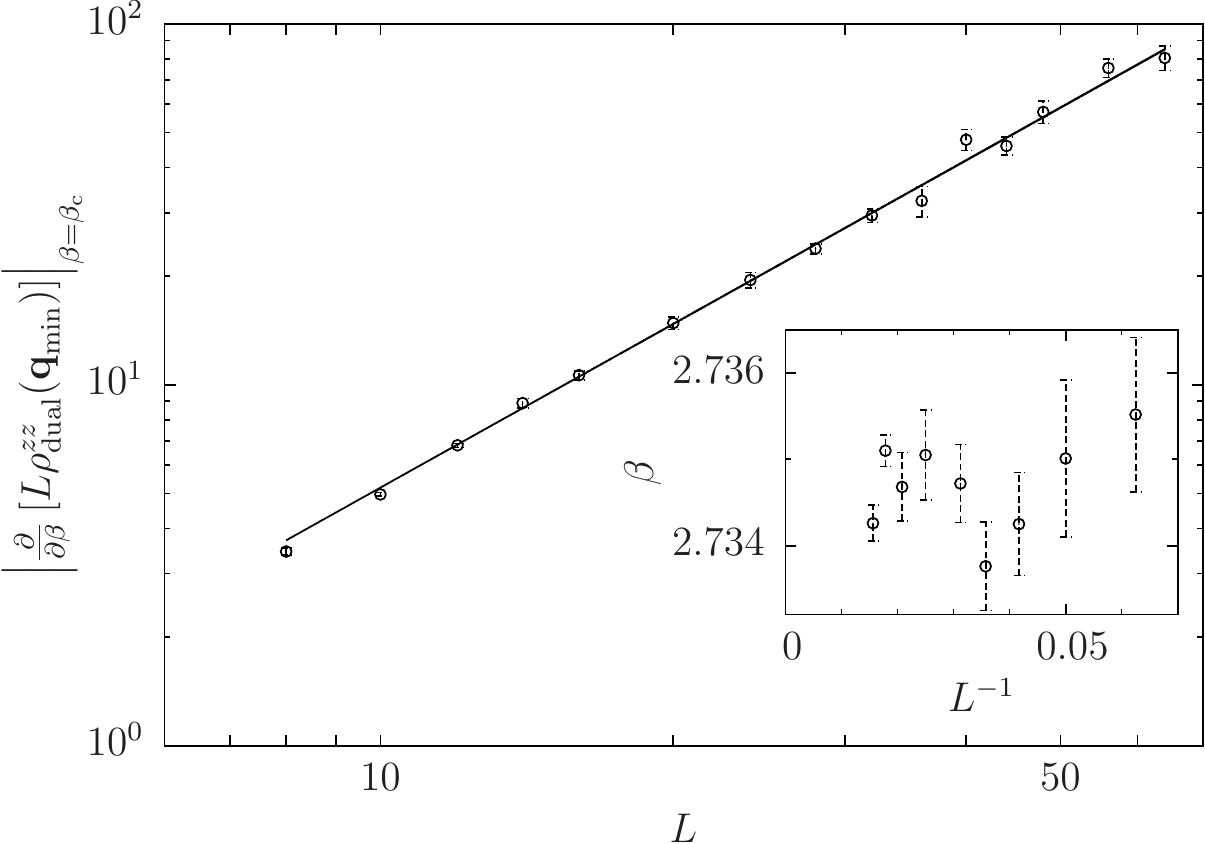}
\caption{Log-log plot of
  $\pderiv{}{\beta}L\rho_{\text{dual}}^{zz}(\v{q}_{\text{min}})$ measured at the critical point
  $\beta_{\text{c}}$, as a function of
  system size $L$. The charge is $e = 5.0$. Measurements are performed
  for 15 different system sizes $L \in \{8, \dots, 64\}$. The
  derivative was found by calculating the differences of
  $\rho_{\text{dual}}^{zz}(\v{q}_{\text{min}})$. The solid straight
  line is the best fit obtained with a fit function on the form $aL^b$
  where $a$ and $b$ are two free parameters. The inset shows the
  $L\rho_{\text{dual}}^{zz}(\v{q}_{\text{min}})$ crossings for systems
  $L$ and $L/2$ as a function of $L^{-1}$. These crossings were used to estimate the critical 
  point, $\beta_{\text{c}} = \num{2.7347 \pm 0.0005}$. 
Errors in determining $\beta_{\text{c}}$ are taken into account by also considering the 
sensitivity of $\nu$ with respect to $\beta$ when 
estimating the uncertainty in the exponent.}
\label{fig:Inv_U1_scaling}
\end{figure}

\subsection{{Estimate for a bicritical point}}
\label{Subsec:Bicritical_region}
In Ref.~\onlinecite{Kuklov_PRL_2008}, the flowgram method has been suggested as a useful tool to assess 
whether or not there is a tricritical point at weak couplings to the gauge field. This method relies on 
resolving a first-order phase transition at stronger couplings, just below the bicritical point at which the 
paired phase opens up between the normal phase and the \groupSU{2} phase. It is thus important to be 
able to determine the bicritical point accurately. For this purpose, we will focus on the region slightly 
\textit{above} the bicritical point and establish when two separate phase transitions are clearly resolved. 
In this way, we can determine an upper bound on the bicritical point.  

\subsubsection{{Signatures of an intermediate paired phase at $e = 4.2$.}}
In order to discern two separate, but close-lying phase transitions, we need to establish signatures that can 
be taken as evidence for splitting of a transition line. To this end, results are presented for the case when 
$e = 4.2$. We find unambiguous evidence for two separate phase transitions. Remarkably, at smaller system sizes we find characteristics of 
the phase transition consistent with a first 
order transition, and it was interpreted as such in Ref.~\onlinecite{Kuklov_PRL_2008}. ($e = 4.2$ corresponds to $g \approx 1.88$ in 
the units of Ref.~\onlinecite{Kuklov_PRL_2008}. This Reference gave the estimate for the position of the bicritical point at $g \approx 2.0$.) 
As we shall see, performing computations on larger systems leads to a different conclusion. The reason is that finite-size effects will 
disguise the existence of separate transitions and make them appear as one.

In Fig.~\ref{fig:Signatures}, results are presented for four different
observables obtained at 12 different system sizes, $L \in \{8, \dots,
56\}$, in a coupling range covering both phase transitions. In panel
(a), results for the specific heat are given. When system sizes are
small, it is only possible to resolve one peak in the specific
heat. However, when $L = 40$, it is possible to resolve a bump to the
left of the peak. The bump, which corresponds to the $\groupO{3}$ ordering phase 
transition, becomes more pronounced when $L$
increases. This behavior suggests that there are two transitions
instead of one. Moreover, in the inset of panel (a) we study the
scaling of the peak on a log-log scale. When $L$ is small, there is a rather steep and
slightly increasing slope. However, at higher values
of $L$ there is a definite change in the slope towards smaller values,
corresponding to a sudden slowing down in the growth of the peak. This
behavior should clearly be associated with resolving separate
transitions with increasing $L$.

\begin{figure}[tbp]
\includegraphics[width=\columnwidth]{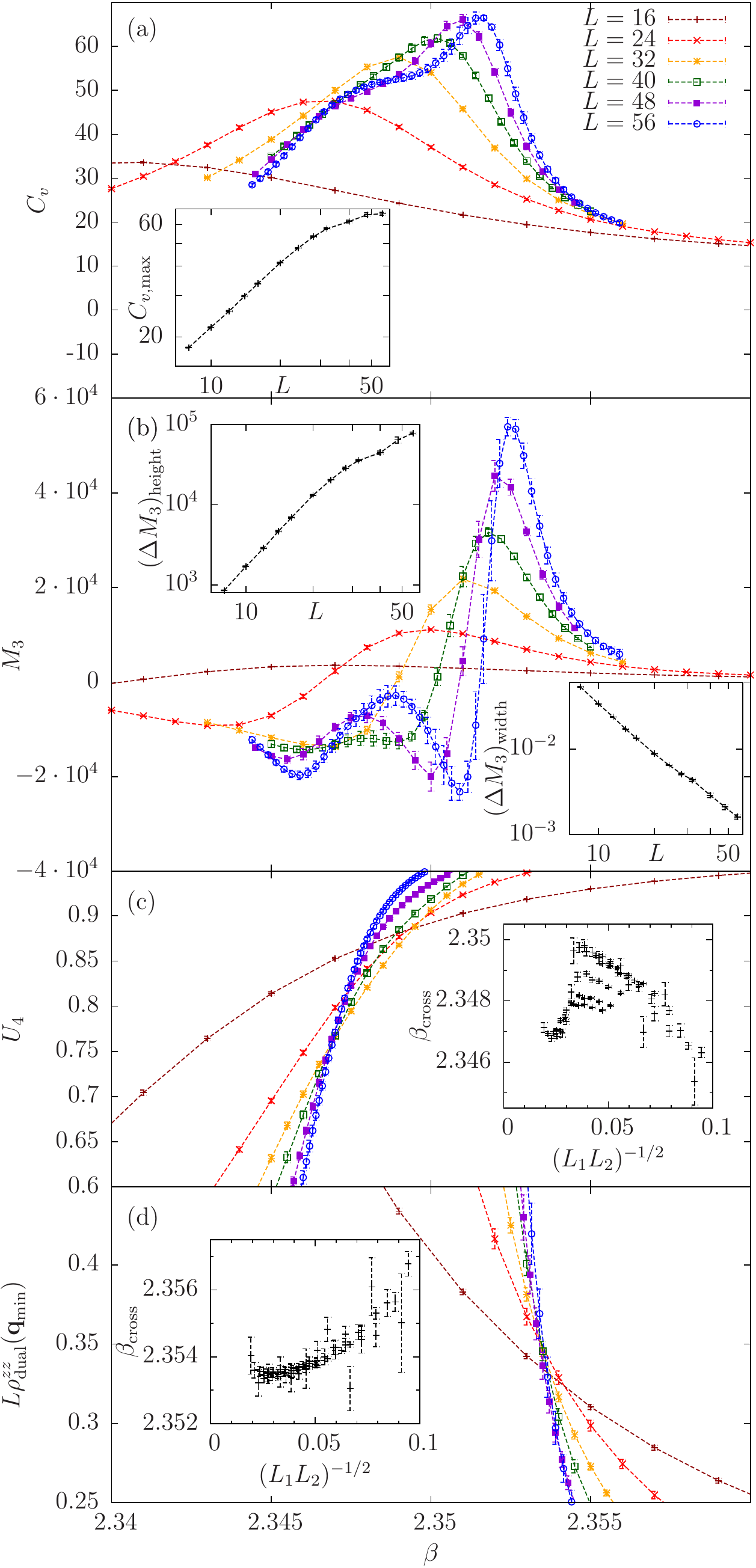}
\caption{(Color online) Monte Carlo results for four different
  quantities and 12 different system sizes obtained for a coupling
  range covering two separate, but close-lying phase transitions. The
  gauge field coupling $e = 4.2$. For clarity, the panels only show results for $L \in \{16, 24, 32, 40, 48,
  56\}$, but insets include all 12 sizes, $L \in \{8, \dots,
  56\}$. Panel (a) shows results for the specific heat $C_v$, and the
  inset shows the scaling of the peak $C_{v, \text{max}}$ in a log-log
  scale. Panel (b) shows the results for the third moment of the action $M_3$, and the
  insets show the scaling of $(\Delta M_3)_{\text{height}}$ and
  $(\Delta M_3)_{\text{width}}$ in a log-log scale. Panel (c) shows the Binder cumulant
  $U_4$, and the inset shows the coupling $\beta_{\text{cross}}$ where the Binder
curves cross as a function of $(L_1 L_2)^{-1/2}$ where $L_1$ and $L_2$
  are the two actual sizes. Panel (d) shows the
  quantity $L\rho_{\text{dual}}^{z z}(\v{q}_{\text{min}})$ and the
  inset shows the coupling where the curves cross. Lines are guide to
  the eyes.
}
\label{fig:Signatures}
\end{figure}

In panel (b) of Fig.~\ref{fig:Signatures}, results for the third
moment of the action are presented. When system sizes are small, it is
only possible to resolve a characteristic form corresponding to a
single phase transition. However, at $L \geq 40$, a secondary form is
developing to the left of the original form, resolving the
$\groupO{3}$ ordering transition. When studying the scaling of the quantities $(\Delta M_3)_{\text{height}}$ and
$(\Delta M_3)_{\text{width}}$ in the insets of the panel, it is clear
that they both exhibit slope changes associated with resolving both transitions.
\footnote{When $L$ is large, such that there are two clearly 
separate transitions, $(\Delta M_3)_{\text{height}}$ and $(\Delta M_3)_{\text{width}}$ are determined by the two extrema of the most prominent 
transition in the $M_3$ plot, which is the Higgs transition.} 

The Binder cumulant is given in panel (c) of Fig.~\ref{fig:Signatures}, and its crossings are given in the inset of the panel. By considering 
the crossings with largest $L$, we find that the critical point of the $\groupO{3}$ ordering transition is $\beta_{\text{c}} = \num{2.347 \pm 0.001}$, 
a value that corresponds well with the leftmost transition point in panel (a) and (b). Note that there is a non-monotonic behavior in the coupling 
values of the Binder crossings. Hence, by studying small systems only, one might be misled to overestimate the critical point of the phase transition.

In panel (d) of Fig.~\ref{fig:Signatures}, we show results for the quantity $L\rho_{\text{dual}}^{z z}(\v{q}_{\text{min}})$, and the corresponding 
crossings are given in the inset. We estimate the critical point of the Higgs transition to be $\beta_{\text{c}} = \num{2.353 \pm 0.001}$ by a 
crude extrapolation to the thermodynamic limit. Hence, the critical point of the Higgs transition is significantly different from the critical 
point of the $\groupO{3}$ ordering transition.

The results in Fig.~\ref{fig:Signatures} show that it is of particular importance to simulate large systems in regions where there might be
multiple phase transitions
in multicomponent gauge theories. Discarding data points for $L > 20$, the crossings in panel (c) and (d) appear to converge to the same
coupling. In panel (a) and (b), we would only resolve a single phase transition with rather strong thermal signatures. 

\subsubsection{Monte Carlo results for $e \in \{3.0, \dots, 4.6\}$}
We first turn our attention to the region with $e < 4.2$ to look for the
signatures that we have established above. Fig.~\ref{fig:Heat_capacity_peak} shows the FSS of the peak in the heat
capacity for $e \in \{3.0, \dots, 4.2\}$. The results show
that there is a definite change in the slope of the scaling of $C_{v,
  \text{max}}$, also for $e = 4.0$ and $3.8$. Note that this signature of
splitting appears at higher $L$ when $e$ is reduced, corresponding to
the coupling difference between the two transitions being smaller. The
slope of the dotted line in Fig.~\ref{fig:Heat_capacity_peak} is the
slope of a first-order transition [see
Eq.~\eqref{eq:specific_heat_fss_first_order}]. For all values of $e$
in Fig.~\ref{fig:Heat_capacity_peak}, we find that for small and
intermediate $L$ the slope is steep and increasing, and one might be
tempted to conclude that they all are first-order transitions. However, the change towards a smaller slope, that we find 
for large $L$ and $e \in \{3.8,4.0, 4.2\}$, is indeed inconsistent with a single first-order phase transition.

\begin{figure}[tbp]
\includegraphics[width=\columnwidth]{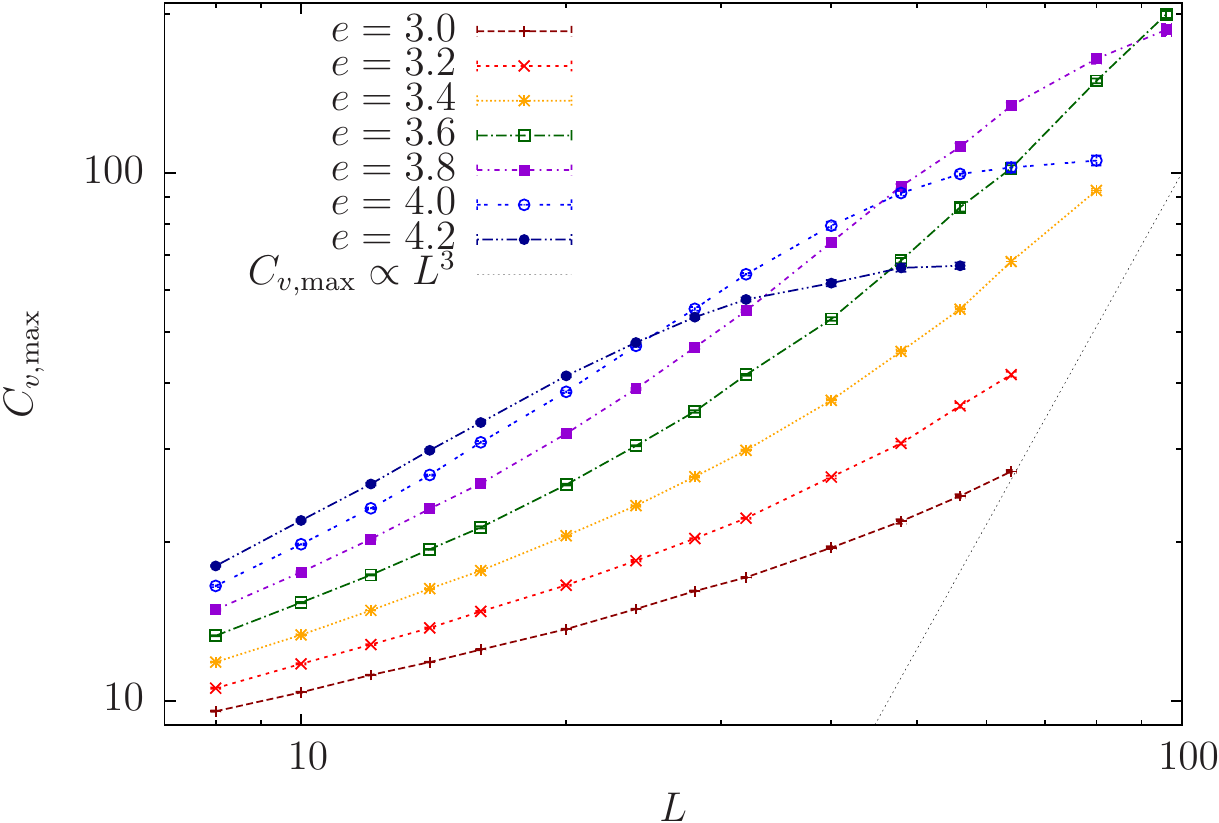}
\caption{(Color online) Log-log plot of the value of the specific heat peak $C_{v, \text{max}}$ as a function 
of system size $L$ for seven different values of $e \in \{3.0, \dots, 4.2\}$. The dotted line corresponds to the 
slope expected for a first-order transition, according to Eq.~\eqref{eq:specific_heat_fss_first_order}. For $e \geq 3.8$, the scaling 
of $C_{v, \text{max}}$ shows a negative curvature, instead of curving up towards the first-order characteristic scaling line. From 
this, our upper bound on the position of the bicritical point in the phase diagram would be $e=3.8$. Lines are 
guide to the eyes.}
\label{fig:Heat_capacity_peak}
\end{figure}

In Fig.~\ref{fig:Third_mom_scaling}, we show the FSS of $(\Delta M_3)_{\text{height}}$ and $(\Delta M_3)_{\text{width}}$. Observe that
the same signatures of splitting appears for $e \in \{3.8, 4.0\}$ as found for $e = 4.2$ above, namely that the slope of $(\Delta
M_3)_{\text{height}}$ changes to a smaller value and the slope of $(\Delta M_3)_{\text{width}}$ changes to a higher value. This is 
again inconsistent with the scaling of a single first-order transition. For a first-order transition the slopes should converge towards the 
scaling for first-order transitions, given in Eqs.~\eqref{eq:third_moment_height_fss_first_order} and
\eqref{eq:third_moment_width_fss_first_order} (see Ref.~\onlinecite{Kragset_PRL_2006} for an example).   

\begin{figure}[tbp]
\includegraphics[width=\columnwidth]{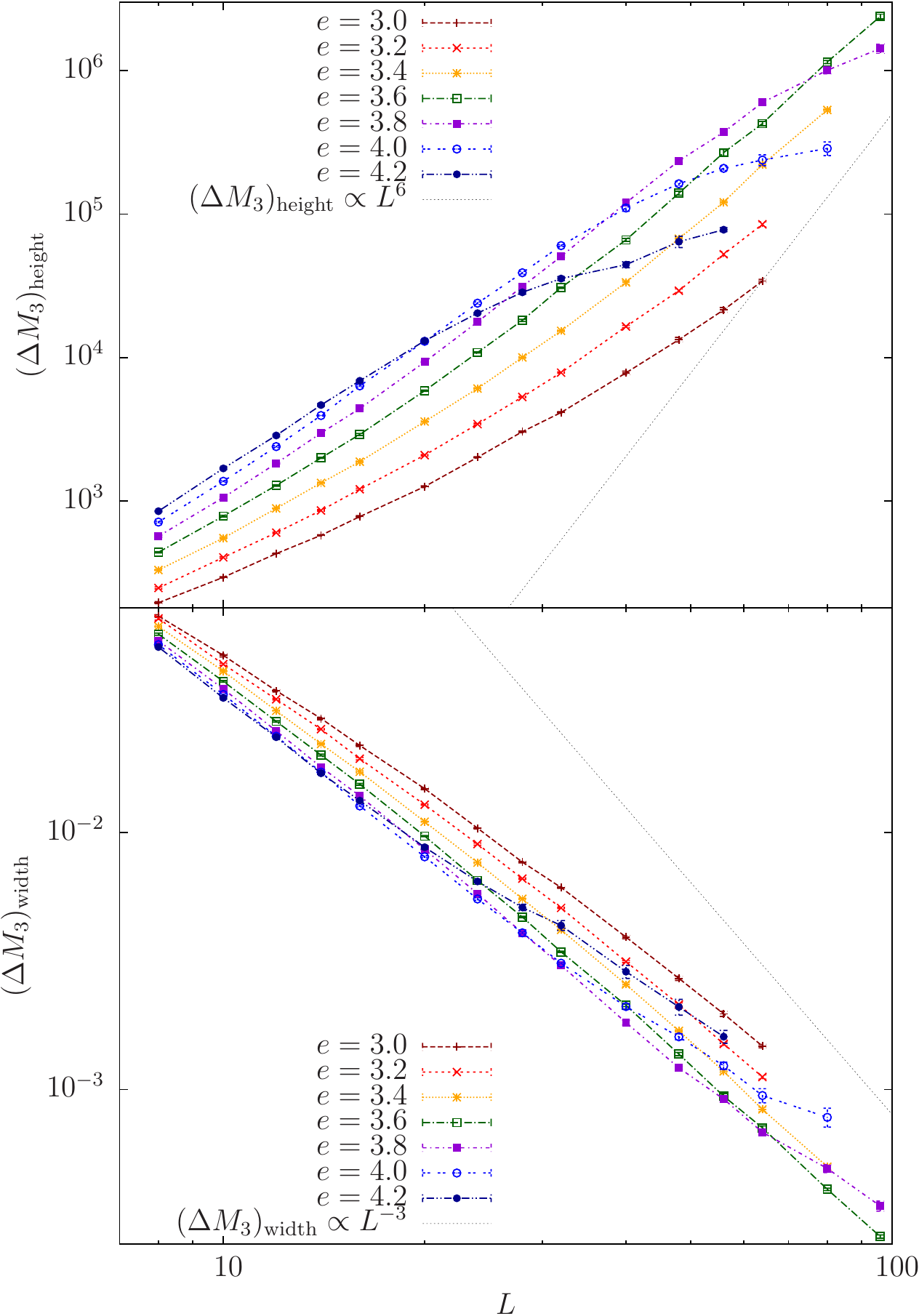}
\caption{(Color online) Log-log plot of the FSS of the height (upper panel) and the width (lower panel) of the third 
moment of the action, for seven different values of $e \in \{3.0, \dots, 4.2\}$. The dotted lines correspond to the 
slope expected for a first-order transition, according to Eqs.~\eqref{eq:third_moment_height_fss_first_order} and 
\eqref{eq:third_moment_width_fss_first_order}. Lines are guide to the eyes.}
\label{fig:Third_mom_scaling}
\end{figure}

To determine the positions  of the $\groupO{3}$ ordering transition and the Higgs transition, the finite size crossings of
$U_4$ and $L\rho_{\text{dual}}^{z z}(\v{q}_{\text{min}})$ are given in Fig.~\ref{fig:Crossings} for eight different values of 
$e \in \{3.2, \dots, 4.6\}$. For $e \in \{4.0, \dots, 4.6\}$, the $U_4$ crossings and the $L\rho_{\text{dual}}^{z z}(\v{q}_{\text{min}})$
crossings clearly extrapolates to different couplings as expected for two separate transitions. Also note the corresponding 
non-monotonic behavior for the Binder crossings. When the coupling difference between the two phase transitions decreases, larger 
systems are needed to resolve this feature. For $e = 3.8$, we observe that the leftmost $U_4$ crossing ($L_1 = 80, L_2 = 96$) deviates, 
consistent with the non-monotonic behavior for the larger $e$ values. For the sizes available, the crossings seem to converge to 
the same coupling value for  $e \in \{3.2, \dots, 3.6\}$.    

\begin{figure}[tbp]
\includegraphics[width=\columnwidth]{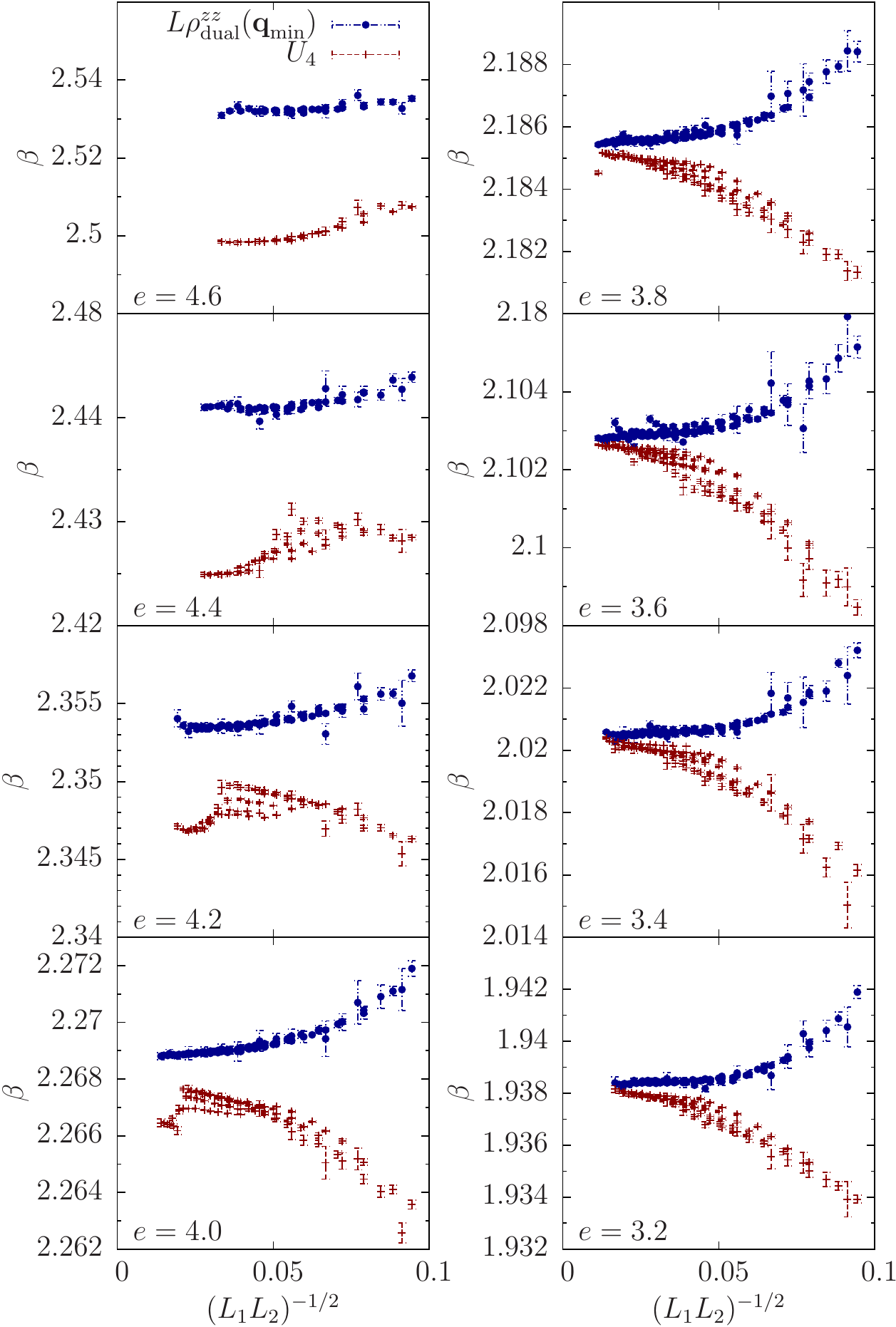}
\caption{(Color online) Plots of the finite size crossings of the Binder cumulant $U_4$ [Eq.~\eqref{eq:O3_Binder_cumulant}], and 
the quantity $L\rho_{\text{dual}}^{z z}(\v{q}_{\text{min}})$ for eight different values of $e \in \{3.2, \dots, 4.6\}$. The x-values are 
given by $(L_1L_2)^{-1/2}$ where $L_1$ and $L_2$ are the two sizes that form the crossing.}
\label{fig:Crossings}
\end{figure}

The results in Figs.~\ref{fig:Heat_capacity_peak}, \ref{fig:Third_mom_scaling} and \ref{fig:Crossings}, show
 that there are two separate transitions when $e \geq 3.8$. 
We thus estimate that the bicritical 
point must be below $e = 3.8$. Clearly, the system sizes we are able to reach are too small to conclusively determine if there are separate 
transitions for $e < 3.8$. However, in order to estimate the bicritical point $e_{\text{bc}}$, in
Fig.~\ref{fig:Difference_critical_coupling} we show results for the coupling difference between the two phase transitions $\Delta\beta_{\text{c}}$ 
as a function of the coupling $e$. To estimate when $\Delta\beta_{\text{c}} \rightarrow 0$, in the lower panel, we show $\Delta\beta_{\text{c}}$
as a function of $e-e^*$ on a log-log scale where $e^*$ is some trial value as labeled in the key of the figure. If $e^* \approx
e_{\text{bc}}$, a straight line should be expected. A positive curvature suggests that $e^* > e_{\text{bc}}$ and a negative curvature
suggests that $e^* < e_{\text{bc}}$. Since there is a clear positive curvature both for $e = 3.8$ and $e = 3.6$, this suggests that
$e_{\text{bc}} < 3.6$. Note that the results given in the lower panel of Fig.~\ref{fig:Difference_critical_coupling} 
essentially is an extrapolation of the difference $\Delta\beta_{\text{c}}$ (which also is an extrapolation) in the upper panel to find the 
point $e_{\text{bc}}$ where $\Delta\beta_{\text{c}} = 0$. As it will be clear below, even at the largest system sizes accessible for us, we could 
not prove that  there is a single first-order transition at  $e = 3.6$. Therefore, simulations of even larger systems are needed to determine more 
accurately the existence and the position of $e_{\text{bc}}$.

\begin{figure}[tbp]
\includegraphics[width=0.9\columnwidth]{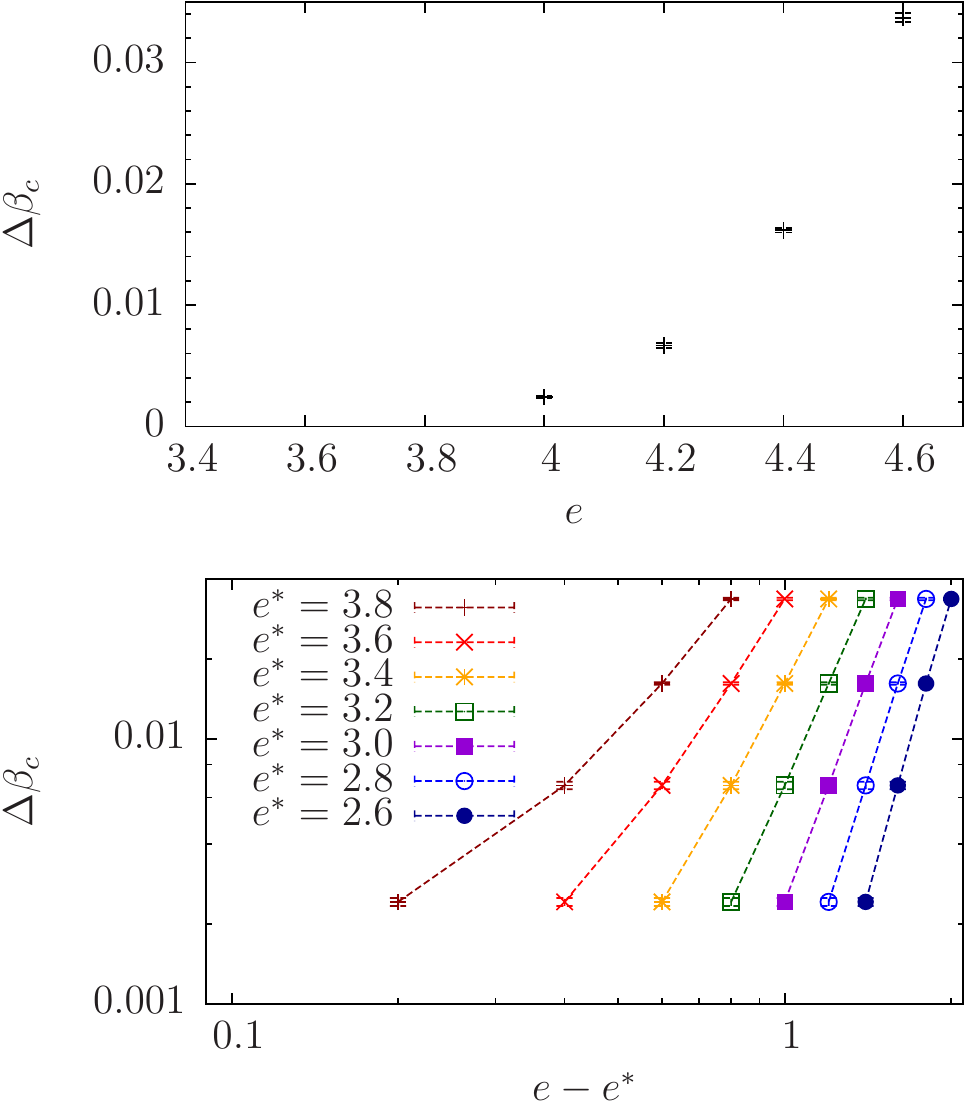}
\caption{(Color online) Plot of the difference in the critical coupling between the
  Higgs transition and the $\groupO{3}$ ordering transition,
  $\Delta\beta_c$. $\Delta\beta_c$ is determined
  by calculating the difference between the $L\rho_{\text{dual}}^{z
    z}(\v{q}_{\text{min}})$ crossing and the $U_4$ crossing, and
  averaging over four of these differences with largest value of
  $(L_1L_2)^{1/2}$ (i.e., the four leftmost data points from the panels
  in Fig.~\ref{fig:Crossings}). We only include results for $e \geq 4.0$ where the
  non-monotonic behavior of the Binder crossings can clearly be
  resolved. Upper panel: $\Delta\beta_c$ as a function of $e$. Lower
  panel: Log-log plot of $\Delta\beta_c$ as a function of $e-e^*$
  where $e^*$ is given in the key. Positive curvature
  suggests that $e^* > e_{\text{bc}}$, negative curvature suggests that $e^*
  < e_{bc}$ and a straight line suggests that $e^* \approx e_{\text{bc}}$. Lines are guide to
  the eyes.}
\label{fig:Difference_critical_coupling}
\end{figure}

Our estimates for the bicritical point differ from the results in Refs.~\onlinecite{Motrunich_ArXiv_2008} and
\onlinecite{Kuklov_PRL_2008} which studied substantially smaller systems. Our \textit{upper} bound $e_{\text{bc}} < 3.8$ corresponds to
$\mathcal{K}_{\text{bc}} > 0.151$ in Ref.~\onlinecite{Motrunich_ArXiv_2008}. This means that a part of the 
line that was interpreted as a direct first-order transition in that work, in fact are two separate transitions. 
Moreover, the upper bound $e_{\text{bc}} < 3.8$ corresponds to $g_{\text{bc}} < 1.65$ in Ref.~\onlinecite{Kuklov_PRL_2008} 
where the bicritical point was estimated to $g \approx 2.0$. 

\subsubsection{Signatures of a weak first-order transition}
Although we are led to a different conclusion concerning the phase diagram than Refs.~\onlinecite{Motrunich_ArXiv_2008} 
and \onlinecite{Kuklov_PRL_2008} for $e \geq 3.8$, we find some of the same thermal signatures. As mentioned above 
(see Figs.~\ref{fig:Heat_capacity_peak} and \ref{fig:Third_mom_scaling}), when systems are too small to resolve two 
phase transitions, the Monte Carlo results show that the scaling of $C_{v, \text{max}}$ and $(\Delta M_3)_{\text{height}}$ 
are almost as one would expect for a single first-order transition. Moreover, when investigating the energy distributions for 
$e \in \{3.8, 4.0\}$ in Fig.~\ref{fig:Histograms}, we find that the histograms are broad. Also, in contrast to previous works, 
we have resolved bimodal structures for $e \in \{3.4, 3.6 \}$. This could be interpreted as evidence of a first order phase 
transition. At the same time we note that  they only appear at the largest system sizes. Thus, 
it is difficult to determine if the correct scaling for first-order transition is obeyed.\cite{Lee_PRL_1990,Lee_PRB_1991} 
The histograms that appear at the largest system sizes, have not yet started to evolve into 
distributions resembling delta functions. In particular, for the system sizes which we can access the dips in between the peaks are 
still increasing with system size, rather than decreasing. The latter is required for drawing a firm conclusion that there is a direct 
first-order phase transition at $e=3.4$ and $e=3.6$. Although rare, there are examples in the literature where bimodal energy distributions 
are found in cases with no first-order phase transition.\cite{Schreiber_J_Phys_A_2005, Behringer_PRE_2006,Fytas_J_Stat_Mech_2008, Jin_PRL_2012}
\begin{figure}[tbp]
\includegraphics[width=\columnwidth]{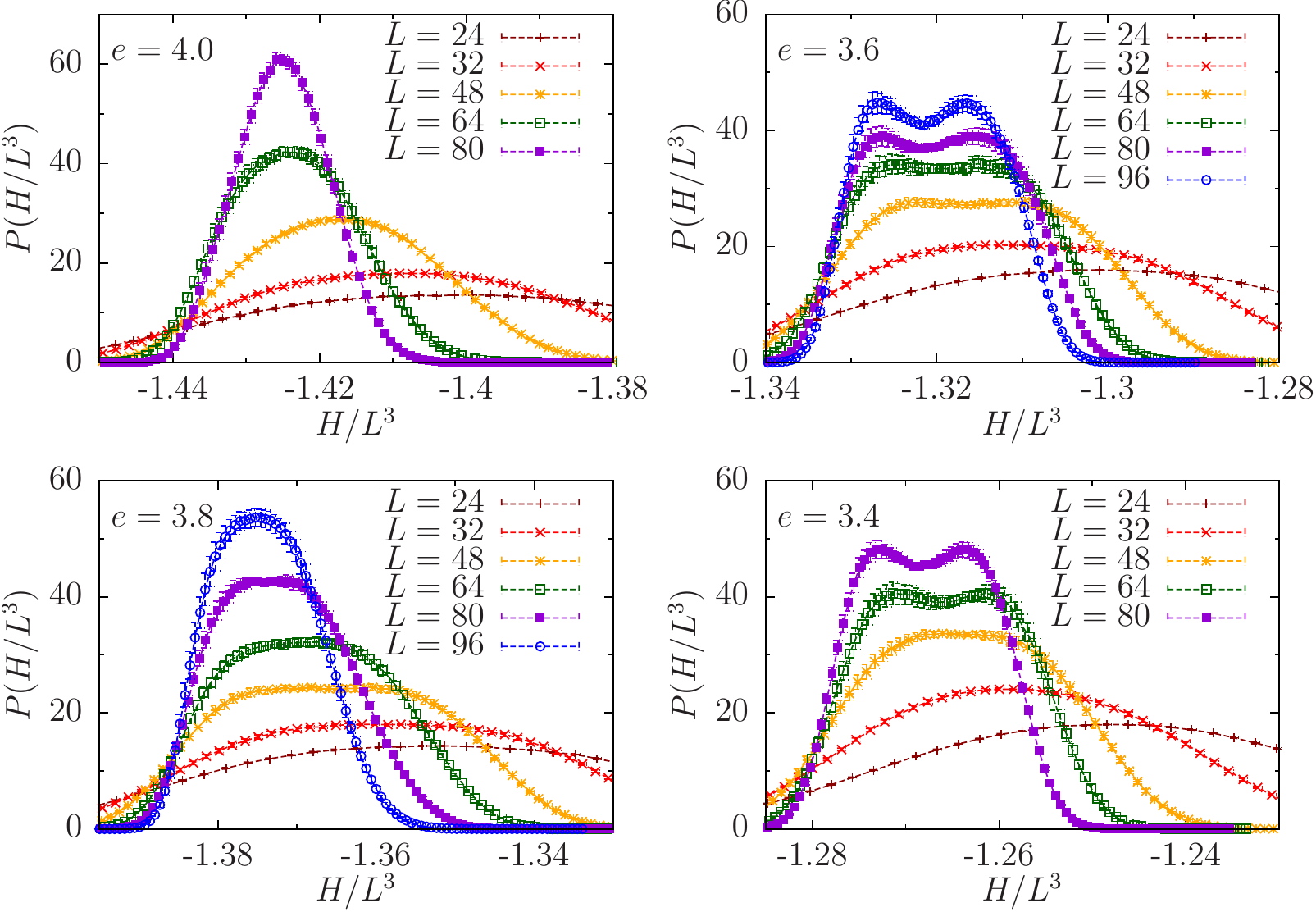}
\caption{(Color online) Histograms of the probability distribution of
  the energy per site $H/L^3$, for $e \in \{3.4, 3.6, 3.8, 4.0\}$. In
  every case the flattest (or most bimodal) energy histograms were
  found by reweighting in the vicinity of the pseudocritical coupling
  corresponding to the peak of the specific heat, $C_{v, \text{max}}$.
  The areas under the curves are normalized to unity.
}
\label{fig:Histograms}
\end{figure}

For $e = 3.8$, we do not resolve any bimodality, but the histograms are wide. The width of the histograms decreases and the flat top structure 
disappears when $L$ increases. This is not consistent with a single first-order transition. Note that if this point is located slightly above 
the bicritical point, then according to a mean-field argument, the Higgs transition should be first-order.\cite{Kuklov_Ann_Phys_2006,Motrunich_ArXiv_2008}
Also, as mentioned above, the instability of composite vortices in type-II \groupSU{2} theory suggests that 
the system should be a type-I superconductor in the proximity of the paired phase (since the paired phase results from proliferation
of composite vortices), {with a possibility of a first order transition via Halperin-Lubensky-Ma mechanism}. We did not consider large 
enough system sizes to resolve this issue.

Combining the results in Figs.~\ref{fig:Heat_capacity_peak}, \ref{fig:Third_mom_scaling} 
and \ref{fig:Histograms}, it appears that for couplings slightly above the estimated bicritical point, 
there are strong thermal signatures in terms of broad energy distributions and rapidly increasing 
peaks in the specific heat and the third moment of the action. However, when system sizes are larger, 
we can explicitly see signatures of splitting for $e \geq 3.8$. We cannot exclude the possibility 
that this may also be the case for some of the couplings with $e < 3.8$. Indeed, the crude extrapolation 
in Fig.~\ref{fig:Difference_critical_coupling} suggests that $e = 3.6$ also is above the bicritical point. 
If so, we should expect to see signatures of splitting for system sizes larger than those available in this 
work. On the other hand, the strong thermal signatures we find for $e < 3.8$ can also
be consistent with a weak single first-order transition. In that case, we should expect to see that proper 
first-order scaling is obeyed for larger system sizes. 
 
Summarizing this part, we find that the strongest signatures for a single first-order phase transition were found at $e = 3.4$ and $e = 3.6$. 
Previous works on smaller systems did not resolve bimodal structure at these couplings. For $e < 3.4$, we did not 
find any bimodal structure in the energy histograms at the system sizes which we can reach.

\subsection{The flowgram method}
\label{Subsec:flowgram}

To analyze situations where it is difficult to resolve and analyze bimodal structures in histograms such as those considered above, 
the authors of Ref.~\onlinecite{Kuklov_Ann_Phys_2006} proposed the flowgram method. By rescaling the linear system size $L \rightarrow C(g)L$, 
where $g = e^2/(4\beta)$ and where $C(g)$ is a monotonous scaling function of the parameter $g$, it may be possible to collapse curves for various 
physical quantities computed at the phase transition, for different system sizes and coupling constants, onto  a single curve.\cite{Kuklov_ArXiv_2008} 
If such uniform scaling is found for all coupling constants, one may conclude that a phase transition has the same characteristics for all these coupling 
constants. For instance, if a first order phase transition were to be found for large coupling constants, and the scaled plots fall on a single line for 
all other coupling constants, one may conclude that the transition is first order for all these coupling constants. To draw such a conclusion, it is very 
important that a broad enough window of systems sizes $L$ is considered, such that there is adequate overlap of datapoints for all coupling constants, 
when the data are plotted in terms of $C(g)L$.

In Fig.~\ref{fig:Flowgram}, we show results of a flowgram analysis of the quantity $L\rho_{\text{dual}}^{z z}(\v{q}_{\text{min}})$ along 
the $\groupO{3}$ ordering transition line. For this analysis, the phase transition is defined to be at the coupling where the Binder 
cumulant $U_4 = 0.775$. With this definition, we will follow the $\groupO{3}$ ordering transition 
line. As mentioned above, $L\rho_{\text{dual}}^{z z}(\v{q}_{\text{min}})$ is a universal quantity for a continuous
Higgs transition, whereas it will diverge $\sim L$ for a first-order transition. We clearly see such diverging behavior 
when $e \geq 3.6$ (not shown here) and the FSS is consistent with $L\rho_{\text{dual}}^{z z}(\v{q}_{\text{min}}) \sim L$. 
In Refs.~\onlinecite{Motrunich_ArXiv_2008} and \onlinecite{Kuklov_ArXiv_2008}, this was interpreted as a first-order 
phase transition. However, a diverging $L\rho_{\text{dual}}^{z z}(\v{q}_{\text{min}})$ is also consistent with being 
above the bicritical point when following the transition line of the $\groupO{3}$ ordering transition. Hence, the 
results in Fig.~\ref{fig:Flowgram} correspond well with there being two closely separated phase transitions for these 
values of $e$, see Figs.~\ref{fig:Signatures}-\ref{fig:Difference_critical_coupling} above.  

\begin{figure}[tbp]
\includegraphics[width=\columnwidth]{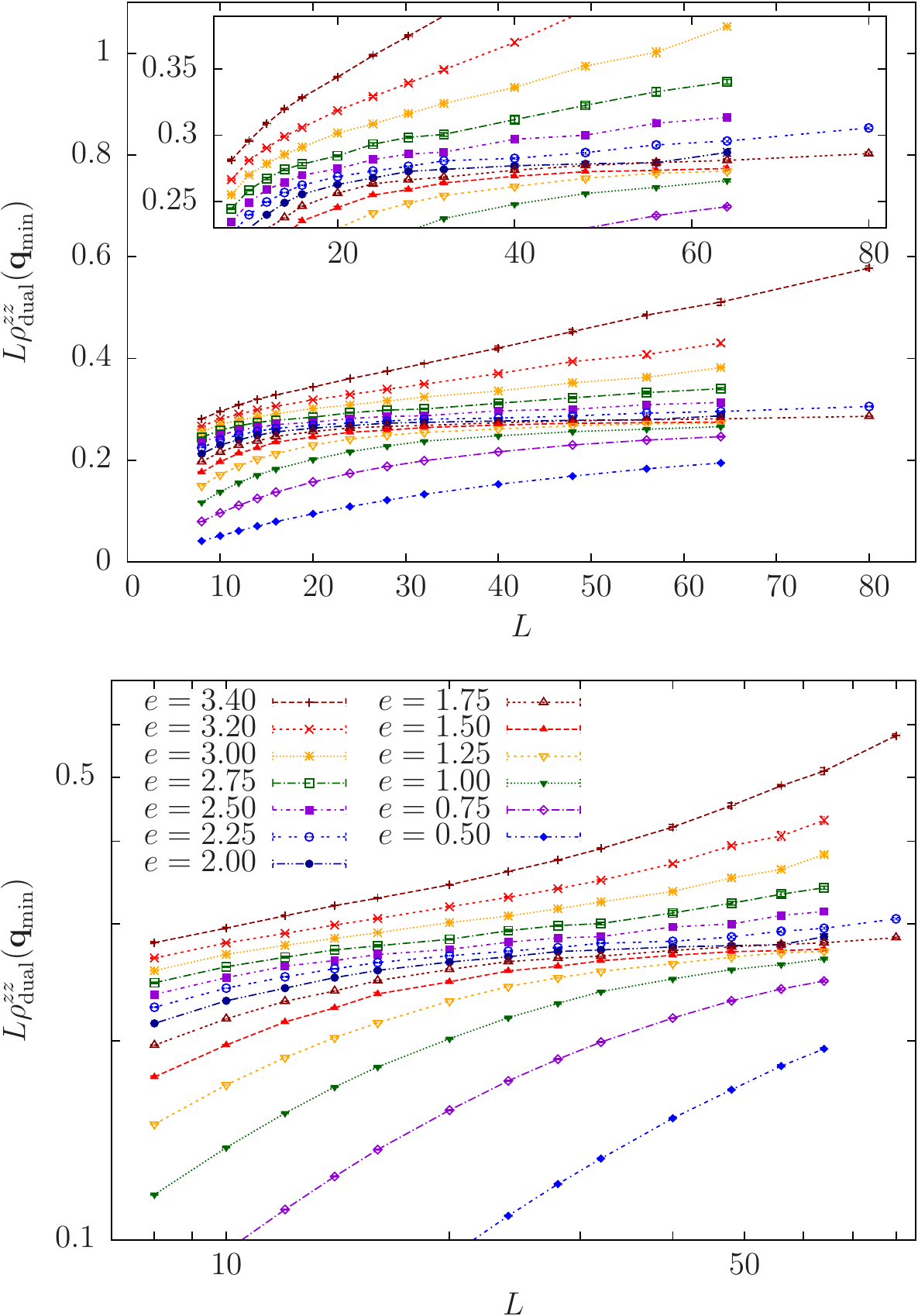}
\caption{(Color online) Flowgram of $L\rho_{\text{dual}}^{z
    z}(\v{q}_{\text{min}})$ along the $\groupO{3}$ ordering
  transition line. In this analysis, the critical point is fixed by
  $U_4 = 0.775$. Then $L\rho_{\text{dual}}^{z
    z}(\v{q}_{\text{min}})$ is measured at this point and plotted as
a function of system size $L$. The results are given for 13 different
values of $e \in \{0.5, \dots, 3.4\}$. The upper panel
shows results on a normal scale and the inset zooms in
on the results for $e \in \{1.0, \dots, 3.0\}$. The lower panel shows the
results on a log-log scale. Lines are guide to the eyes.}
\label{fig:Flowgram}
\end{figure}

For $e \in \{3.0, 3.2, 3.4\}$, the flowgram analysis suggests that $L\rho_{\text{dual}}^{zz}(\v{q}_{\text{min}})$ diverges, but the
FSS is weaker than $\sim L$ for the sizes available.  This is consistent with either being above the bicritical point, or with a 
first-order transition. {For smaller couplings, the large size behavior of the flowgrams is hard to determine. In particular, for 
the couplings $e \leq 2.0$ the flowgrams seem to converge slowly to a fixed value, but one cannot rule out diverging behavior at larger sizes.}

In Fig.~\ref{fig:Rescaled_flowgram}, we plot the results for the flowgram data in Fig.~\ref{fig:Flowgram} in terms
of the variable $C(g)L$ on a log-log scale, using the scaling function $C(g) = 3.0324 g + 0.0997[\exp(4.1005 g)-1]$.\footnote{This is not the 
same scaling function as suggested in Ref.~\onlinecite{Kuklov_ArXiv_2008}. In that work, the system sizes were
smaller than in this work. Because of finite-size effects, the best scaling function may change slightly when 
larger systems are included. In this context, the best scaling function is determined by requiring the best
collapse for  the largest system sizes.} For large values of $C(g)L$, the  
collapse appears to be good, and consistent with Ref.~\onlinecite{Kuklov_ArXiv_2008}.
In our case, we note that for various couplings there are sizeable finite-size effects which make it impossible to collapse 
smaller systems onto the same master curve. Removing the data points for the smallest systems for each coupling constant 
would improve the collapse considerably.

\begin{figure}[tbp]
\includegraphics[width=\columnwidth]{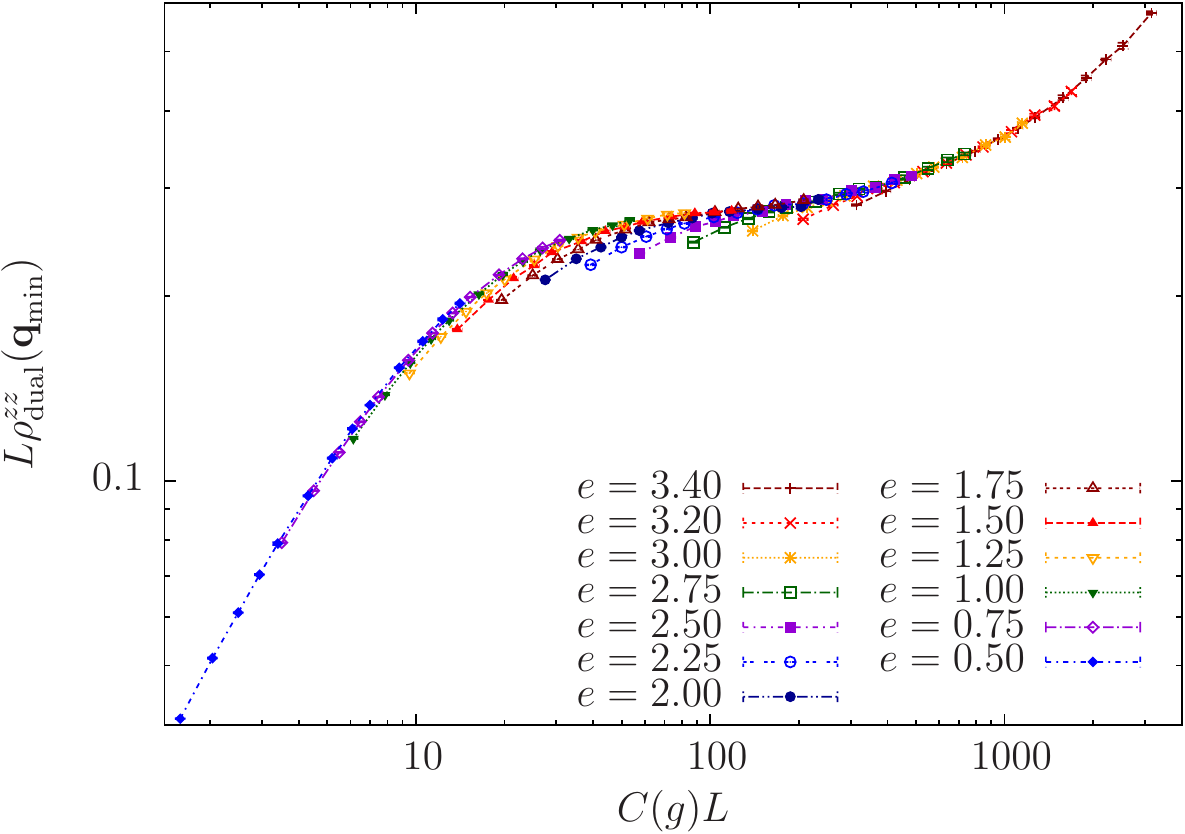}
\caption{(Color online) Rescaled flowgram of the data in
  Fig.~\ref{fig:Flowgram}. 
  The system size $L$ is rescaled by $L
  \rightarrow C(g)L$ where $C(g) = 3.0324 g + 0.0997[\exp(4.1005 g)-1]$
  with $g = e^2/(4\beta)$.  Lines are guide to the eyes.}
\label{fig:Rescaled_flowgram}
\end{figure}

What can the results of Figs. \ref{fig:Flowgram} and \ref{fig:Rescaled_flowgram} tell us about the character of the phase transition, and about 
the existence of a possible tricritical point separating a line of first order phase transitions from a critical line? In Fig. \ref{fig:Rescaled_flowgram}, 
the presence of a tricritical point and a line of second order phase transition would show up as a bifurcations of the master curve at 
large $C(g) L$. In Ref. ~\onlinecite{Kuklov_Ann_Phys_2006}, a tricritical point in a global $\groupU{1} \times \groupU{1}$ model was detected via a breakdown of the 
curve collapse just below a tricritical point. We did not observe such a breakdown of the curve collapse for the $\text{NCCP}^1$ model. There may exist special 
cases where the universalities of the line of second order phase transitions and of a tricritical endpoint are quite similar. Then, one may not be able to 
resolve  different plateaus at finite system sizes. In such a situation for large couplings $e$, we would have the behavior shown in 
Fig. \ref{fig:Rescaled_flowgram}. For small couplings, there should appear another horizontal branch of the scaling function at large values of the 
argument $C(g) L$, were a tricritical point to exist. The results in  Fig. \ref{fig:Rescaled_flowgram} show no such feature. However, note that the data 
points for $e \leq 1.50$ only extend to about the middle of the plateau in Fig. \ref{fig:Rescaled_flowgram}. This illustrates the fact, which is also 
obvious from the lower panel of Fig. \ref{fig:Flowgram}, that for small couplings $e \leq 1.50$, we have not reached large enough system sizes to be able
to ascertain if the curves are horizontal, or if there is an upward curvature in any of the curves for $e \leq 1.50$. Consider for instance 
the coupling $e=1.50$, which is the curve in Fig. \ref{fig:Flowgram} which features the most pronounced horizontal part for the system 
sizes we have studied. In Fig. \ref{fig:Rescaled_flowgram}, this curve extends out to $C(g) L \approx 120$, which is in the middle of 
the plateau. To ascertain whether this curve falls on the upward curving master curve or continues horizontally would require
an extension of the curve out to $C(g) L \approx 400$, or system sizes of about $300^3$. No computations 
have been performed on these types of systems remotely approaching this range. Another way of in principle detecting a tricritical point would be as
follows. Suppose that one, in order to get good data collapse for the entire range of coupling constants would need to resort to two
different types of scaling functions, one below some coupling constant and another one above this coupling constant. At the
point where these functions are joined, one typically has a non-analyticity. One can thus in principal locate a tricritical point at $g=g_{\text{tri}}$ by 
detecting a non-analyticity in  $C(g)$.\cite{Kuklov_pcomm_2012} With our current data we have not resolved such a feature in $C(g)$.

\section{Summary}
In this work, we have studied the three dimensional \groupSU{2}-symmetric noncompact $\text{CP}^1$ model.
We have implemented an algorithm which permits us to perform an investigation of the model at substantially larger system sizes than those reached in 
previous works. It has been shown that at couplings $e=3.8$ and $e=4.0$, which were previously estimated to belong to the regime where the system undergoes a single 
first-order phase transition, certain signatures should be taken as direct evidence of two separate phase transitions. {Hence, 
we conclude that a bicritical point must be located below $e = 3.8$.} 
We find  bimodality in histograms, consistent with early stages of development of a first order transition, at $e=3.4$ and $e=3.6$ 
(though the histograms do not yet resemble two $\delta$-functions and thus indeed it cannot represent
a proof of a first order phase transition\cite{Kuklov_APS_2008,Motrunich_ArXiv_2008, Kuklov_PRL_2008} )
\footnote{In Ref.~\onlinecite{Kuklov_PRL_2008}, a bimodal distribution was seen for $e=3.8$, where we do not observe bimodality. 
Ref.~\onlinecite{Kuklov_PRL_2008}, however,  finds bimodality in other quantities than we consider. We have evidence that $e=3.8$ 
is above the bicritical point. A previously discussed scenario is that there are also first-order transitions above the bicritical point.
}. 
Although our estimate for the position of bicritical point is different,
the data collapse which we find is overall consistent with 
Ref.~\onlinecite{Kuklov_PRL_2008}. 

\acknowledgments
We acknowledge useful discussions with A. Kuklov, F. S. Nogueira, N.~V. Prokof'ev, A. W. Sandvik, B.~V. Svistunov and I.~B. Sperstad. 
E.~V.~H. and T.~A.~B. thank NTNU for financial support. E.~B., and A.~S. thank the Aspen Center for Physics for hospitality and support 
under the NSF grant $\#1066293$. The work was also supported through the Norwegian consortium for high-performance computing (NOTUR). 
AS was supported through the Research Council of Norway, through Grants 205591/V20 and 216700/F20. 
E.~B. was supported by US National Science Foundation CAREER Award No. DMR-0955902, and by the Knut and Alice Wallenberg Foundation 
through the Royal Swedish Academy of Sciences, Swedish
Research Council.
\appendix 
\section{Mapping the $\text{NCCP}^1$ model to a $J$-current model} \label{app:J_current_formulation}
We start with the lattice formulation of the $\text{NCCP}^1$ model,
\begin{gather}
 Z = \prod_{c,\v{r}} \int \diff \psi_{c,\v{r}} \diff \psi_{c,\v{r}}^* \prod_{\mu,\v{r}} \int \diff A_{\mu,\v{r}} \ \e{-S}, \label{eq:A_partition_function}\\
 S = S_t + S_g, \\
 S_t \equiv -t \sum_{c,\mu,\v{r}} \psi_{c,\v{r}} \psi_{c,\v{r}+\uv{\mu}}^* \e{\i A_{\mu,\v{r}}} + \text{c.c.} \label{eq:A_S_t}, \\
 S_g \equiv \frac{1}{8g} \sum_{\mu,\v{r}}\left(\sum_{\nu, \lambda} \epsilon_{\mu\nu\lambda}\Delta_{\nu}A_{\lambda,\v{r}}\right)^2, \\
 \abs{\psi_{1,\v{r}}}^2 + \abs{\psi_{2,\v{r}}}^2= 1 \quad \forall \v{r}, \label{eq:A_CP1_constraint}
\end{gather}
where we have introduced $t \equiv \beta/2$ and $g \equiv e^2/(4\beta)$ -- the same coupling constants as in Ref.~\onlinecite{Kuklov_PRL_2008}. Writing the complex 
fields on polar form,
\begin{gather}
 \psi_{c,\v{r}} = \rho_{c,\v{r}}\e{\i \theta_{c,\v{r}}},\\
 \int \diff \psi_{c,\v{r}} \diff \psi_{c,\v{r}}^* = \int_{0}^{2\cpi} \diff \theta_{c,\v{r}} \int_{0}^{\infty}\rho_{c,\v{r}}\diff \rho_{c,\v{r}},
\end{gather}
we note that the constraint \eqref{eq:A_CP1_constraint} becomes
\begin{equation}
 {\rho_{1,\v{r}}}^2 + {\rho_{2,\v{r}}}^2 = 1, \quad \forall \v{r},
\end{equation}
which describes the unit circle in the first quadrant of the $\rho_{1,\v{r}}\rho_{2,\v{r}}$-plane (since $\rho_{c,\v{r}} \geq 0$). This means that we can incorporate the constraint directly into the integral by introducing the new field $\phi$,
\begin{gather}
 \rho_{1,\v{r}} = \cos \phi_{\v{r}}, \quad  \rho_{2,\v{r}} = \sin \phi_{\v{r}} \\
 \left.\prod_c \int_0^\infty \rho_{c,\v{r}} \diff \rho_{c,\v{r}}\right|_{\sum_c {\rho_{c,\v{r}}}^2 = 1} = \int_{0}^{\frac{\cpi}{2}} \cos \phi_{\v{r}} \sin \phi_{\v{r}} \diff \phi_{\v{r}},
\end{gather}
such that \eqref{eq:A_partition_function}, \eqref{eq:A_S_t} and \eqref{eq:A_CP1_constraint} can be replaced by
\begin{gather}
 Z = \prod_{\v{r}} \int_{0}^{2\cpi} \diff \theta_{1,\v{r}}\diff \theta_{2,\v{r}} \int_{0}^{\frac{\cpi}{2}} \cos \phi_{\v{r}} \sin \phi_{\v{r}} \diff \phi_{\v{r}} \prod_{\mu,\v{r}} \int \diff A_{\mu,\v{r}} \ \e{-S}, \label{eq:A_new_partition_function}\\
 \begin{aligned}
    S_t = -t \sum_{\mu,\v{r}} \left[\cos \phi_{\v{r}} \cos \phi_{\v{r}+\uv{\mu}}\left(\e{\i\left(\theta_{1,\v{r}} - \theta_{1,\v{r} + \uv{\mu}} + A_{\mu,\v{r}} \right)} + \text{c.c.}\right)\right. \\
    + \left.\sin \phi_{\v{r}} \sin \phi_{\v{r}+\uv{\mu}}\left(\e{\i\left(\theta_{2,\v{r}} - \theta_{2,\v{r} + \uv{\mu}} + A_{\mu,\v{r}} \right)} + \text{c.c.}\right)\right].
    \label{eq:A_S_t_new}
 \end{aligned}
\end{gather}

Next, we focus on the the $\theta$-dependent part of the integrand, namely $\exp(-S_t)$, aiming at replacing this field with a $J$-current field. First we symmetrize \eqref{eq:A_S_t_new}: Assuming periodic boundary conditions and using that 
\begin{equation}
 A_{\mu,\v{r}-\uv{\mu}} = -A_{-\mu,\v{r}},
 \label{eq:A_gauge_field_symmetry}
\end{equation}
we get
\begin{multline}
S_t = -\frac{t}{2} \sum_{\kappa,\v{r}} \left[\cos \phi_{\v{r}} \cos \phi_{\v{r}+\uv{\kappa}}\left(\e{\i\left(\theta_{1,\v{r}} - \theta_{1,\v{r} + \uv{\kappa}} + A_{\kappa,\v{r}} \right)} + \text{c.c.}\right)\right. \\
    + \left.\sin \phi_{\v{r}} \sin \phi_{\v{r}+\uv{\kappa}}\left(\e{\i\left(\theta_{2,\v{r}} - \theta_{2,\v{r} + \uv{\kappa}} + A_{\kappa,\v{r}} \right)} + \text{c.c.}\right)\right],
\end{multline}
where $\kappa$ runs over negative as well as positive lattice directions, $\kappa \in \{\pm x, \pm y,\pm z\}$. Then we split $\exp(-S_t)$ into its individual factors and Taylor expand each of them:
\begin{gather}
 \begin{split}
 \e{-S_t} = &\prod_{\kappa,\v{r}} \sum_{\substack{k_{1,\kappa,\v{r}} = 0\\ l_{1,\kappa,\v{r}} = 0}}^{\infty} \sum_{\substack{k_{2,\kappa,\v{r}} = 0\\ l_{2,\kappa,\v{r}} = 0}}^{\infty} \\
 \Bigg[ &\frac{\left(\frac{t}{2} \cos \phi_{\v{r}} \cos \phi_{\v{r}+\uv{\kappa}}\right)^{k_{1,\kappa,\v{r}} + l_{1,\kappa,\v{r}}}}{k_{1,\kappa,\v{r}}!l_{1,\kappa,\v{r}}!} \times \\
 &\frac{\left(\frac{t}{2} \sin \phi_{\v{r}} \sin \phi_{\v{r}+\uv{\kappa}}\right)^{k_{2,\kappa,\v{r}} + l_{2,\kappa,\v{r}}}}{k_{2,\kappa,\v{r}}!l_{2,\kappa,\v{r}}!} \times \\
 &\e{\i(k_{1,\kappa,\v{r}} - l_{1,\kappa,\v{r}})\left(\theta_{1,\v{r}} - \theta_{1,\v{r} + \uv{\kappa}} + A_{\kappa,\v{r}} \right)} \times \\
 &\e{\i(k_{2,\kappa,\v{r}} - l_{2,\kappa,\v{r}})\left(\theta_{2,\v{r}} - \theta_{2,\v{r} + \uv{\kappa}} + A_{\kappa,\v{r}} \right)} \Bigg]
 \end{split} \label{eq:A_Taylor_expansion}
\end{gather}
The factors of the product over the lattice and directions in \eqref{eq:A_Taylor_expansion} may be rearranged such that all the terms containing $\theta_{c,\v{r}}$ are collected into one,
\begin{multline}
 \e{-S_t} = \sum_{\{k,l\}} \prod_{c,\v{r}} \e{\i\theta_{c,\v{r}}\sum_\kappa \left(k_{c,\kappa,\v{r}} - l_{c,\kappa,\v{r}} - k_{c,\kappa,\v{r} - \uv{\kappa}} + l_{c,\kappa,\v{r} - \uv{\kappa}}\right)} \\ \times  \text{(Everything else)}.
\end{multline}
Here $\{k,l\}$ denotes the set of all possible Taylor expansion index field configurations. Inserting this in the partition function \eqref{eq:A_new_partition_function}, the $\theta$-integrals may now be performed. The result is Dirac delta functions (up to an irrelevant scaling factor, which we ignore) at each lattice point, revealing the (\enquote{$J$-current}) constraint
\begin{equation}
 \sum_\kappa k_{c,\kappa,\v{r}} - l_{c,\kappa,\v{r}} - k_{c,\kappa,\v{r} - \uv{\kappa}} + l_{c,\kappa,\v{r} - \uv{\kappa}} = 0, \quad \forall c,\v{r}.
 \label{eq:A_J_current_constraint}
\end{equation}

It is convenient to introduce the non-negative bond subcurrents
\begin{equation}
 J_{c,\kappa,\v{r}} \equiv k_{c,\kappa,\v{r}} + l_{c,-\kappa,\v{r} + \uv{\kappa}} \in \mathbb{N}_0,
 \label{eq:A_J_current_def}
\end{equation}
as well as the total bond currents
\begin{equation}
  \quad I_{c,\kappa,\v{r}} \equiv J_{c,\kappa,\v{r}} - J_{c,-\kappa,\v{r} + \uv{\kappa}} \in \mathbb{Z}.
  \label{eq:A_I_current_def}
\end{equation}
Reordering the sum, the constraint \eqref{eq:A_J_current_constraint} then simplifies to
\begin{equation}
 \sum_\kappa I_{c,\kappa,\v{r}} = 0, \quad \forall c,\v{r};
 \label{eq:A_I_constraint}
\end{equation}
the current conservation in each component at each lattice site.

Getting rid of the $\theta$-field we turn our attention to the $\phi$-field. The terms containing $\phi_{\v{r}}$ for a given $\v{r}$ are on the form
\begin{multline}
 \int_{0}^{\frac{\cpi}{2}} \diff \phi_{\v{r}} \cos^{1 + 2\mathcal{N}_{1,\v{r}}} \phi_{\v{r}} \sin^{1 + 2\mathcal{N}_{2,\v{r}}} \phi_{\v{r}} \times \text{(Everything else)}
 \\ = \frac{\mathcal{N}_{1,\v{r}}!\mathcal{N}_{2,\v{r}}!}{2\left(\mathcal{N}_{1,\v{r}} + \mathcal{N}_{2,\v{r}} + 1\right)!} \times \text{(Everything else)},
 \label{eq:A_phi_integration}
\end{multline}
where, using \eqref{eq:A_J_current_def},\eqref{eq:A_I_current_def} and \eqref{eq:A_I_constraint},
\begin{align}
 \mathcal{N}_{c,\v{r}} &\equiv \frac{1}{2}\sum_\kappa k_{c,\kappa,\v{r}} + l_{c,\kappa,\v{r}} + k_{c,\kappa,\v{r}-\uv{\kappa}} + l_{c,\kappa,\v{r}-\uv{\kappa}} \nonumber \\
 &= \frac{1}{2} \sum_\kappa J_{c,\kappa,\v{r}} + J_{c,-\kappa,\v{r} + \uv{\kappa}} \nonumber \\
 &= \sum_\kappa J_{c,\kappa,\v{r}} \in \mathbb{N}_{0}.
\end{align}

The Taylor expansion \eqref{eq:A_Taylor_expansion} contains an index field dependent factor as well,
\begin{equation}
 \sum_{\{k,l\}}\prod_{c,\kappa,\v{r}} \frac{\left(\frac{t}{2}\right)^{k_{c,\kappa,\v{r}} + l_{c,\kappa,\v{r}} }}{k_{c,\kappa,\v{r}}! l_{c,\kappa,\v{r}}!},
 \label{eq:A_index_field_part}
\end{equation}
which we want to write as a function of the $J$-subcurrent field instead. It is easy to see that
\begin{equation}
 \prod_{c,\kappa,\v{r}} \left(\frac{t}{2}\right)^{k_{c,\kappa,\v{r}} + l_{c,\kappa,\v{r}}} = \prod_{c,\kappa,\v{r}} \left(\frac{t}{2}\right)^{J_{c,\kappa,\v{r}}}
 \label{eq:A_index_field_1}
\end{equation}
by reordering the terms in the product. Using the definition \eqref{eq:A_J_current_def}, as well as some standard combinatorial results, we may rewrite the denominator part of \eqref{eq:A_index_field_part} as
\begin{align}
 \sum_{\{k,l\}}\prod_{c,\kappa,\v{r}} \frac{1}{k_{c,\kappa,\v{r}}! l_{c,\kappa,\v{r}}!} &= \sum_{\{J\}}\prod_{c,\kappa,\v{r}} \sum_{k_{c,\kappa,\v{r}} = 0}^{J_{c,\kappa,\v{r}}}\frac{1}{k_{c,\kappa,\v{r}}! (J_{c,\kappa,\v{r}} - k_{c,\kappa,\v{r}})!}  \nonumber \\
 &= \sum_{\{J\}}\prod_{c,\kappa,\v{r}} \frac{1}{J_{c,\kappa,\v{r}}!} \sum_{k_{c,\kappa,\v{r}} = 0}^{J_{c,\kappa,\v{r}}} \binom{J_{c,\kappa,\v{r}}}{k_{c,\kappa,\v{r}}} \nonumber \\
 &= \sum_{\{J\}}\prod_{c,\kappa,\v{r}} \frac{2^{J_{c,\kappa,\v{r}}}}{J_{c,\kappa,\v{r}}!},
 \label{eq:A_index_field_2}
\end{align}
where $\{J\}$ denotes the set of all possible subcurrent configurations. (There is no problem in summing $k$ away, as it is an independent variable, and all other terms in the partition function are exclusively $J$-dependent -- as we will see in a moment.) Inserting \eqref{eq:A_index_field_1} and \eqref{eq:A_index_field_2} into \eqref{eq:A_index_field_part} gives
\begin{equation}
 \sum_{\{J\}}\prod_{c,\kappa,\v{r}} \frac{t^{J_{c,\kappa,\v{r}}}}{J_{c,\kappa,\v{r}}!},
 \label{eq:A_index_field_rewritten}
\end{equation}
which is what we desired.

Lastly, we want to integrate out the gauge field. The gauge field dependent factors of \eqref{eq:A_Taylor_expansion} are on the form
\begin{equation}
\exp\left[\i\sum_{c,\kappa,\v{r}} A_{\kappa, \v{r}} \left(k_{c,\kappa, \v{r}} - l_{c,\kappa, \v{r}} \right)\right] = \exp\left[\i\sum_{c,\mu,\v{r}} A_{\mu, \v{r}} I_{c,\mu,\v{r}}\right].
\label{eq:A_S_t_gauge_field_contribution}
\end{equation}
Note that the summation is over \emph{only positive} directions on the RHS. (The RHS is found by expanding and reordering the sum in the exponent on the LHS and applying the identity \eqref{eq:A_gauge_field_symmetry} and the bond current definition \eqref{eq:A_I_current_def}.) Combining \eqref{eq:A_S_t_gauge_field_contribution} with $\exp(-S_g)$, the total gauge field contribution to the partition function reads (up to an irrelevant scaling factor)
\begin{multline}
 \prod_{\mu,\v{r}} \int \diff A_{\v{r}} \ \exp\sum_{\mu,\v{r}} \Bigg[\i A_{\mu, \v{r}} \left(I_{1,\mu,\v{r}} + I_{2,\mu,\v{r}}\right) \\ - (8g)\inv \left(\sum_{\nu, \lambda} \epsilon_{\mu\nu\lambda}\Delta_{\nu}A_{\lambda,\v{r}}\right)^2\Bigg] \\
 \propto \exp\Bigg(-\frac{g}{2}\sum_{\substack{c,c', \\ \mu, \v{r},\v{r}'}} I_{c,\mu,\v{r}} V_{\v{r},\v{r}'} I_{c',\mu,\v{r}'} \Bigg),
 \label{eq:A_gauge_field_part}
\end{multline}
where we have applied the Coulomb gauge $\Delta_{\mu} A_{\mu,\v{r}} = 0$. $V_{\v{r},\v{r}'}$ is a long range potential given by by the inverse Fourier transform
\begin{equation}
 V_{\v{r},\v{r}'} \equiv \mathcal{F}\inv \left\{\left[\sum_{\mu} \sin^2\left(\frac{q_\mu}{2}\right)\right]\inv \right\}\left(\v{r} - \v{r}' \right),
\end{equation}
where $q_\mu$ is the $\mu$ component of the Fourier space wave vector $\v{q}$.

Combining everything, \eqref{eq:A_I_constraint}, \eqref{eq:A_phi_integration}, \eqref{eq:A_index_field_rewritten}, and \eqref{eq:A_gauge_field_part}, leaving out trivial scaling factors, we end up with
\begin{widetext}
\begin{equation}
 Z = \sum_{\{J|\sum_\kappa I_\kappa = 0\}} \left[\prod_{c,\kappa,\v{r}} \frac{t^{J_{c,\kappa,\v{r}}}}{J_{c,\kappa,\v{r}}!}\right]
 \left[\prod_{\v{r}} \frac{\mathcal{N}_{1,\v{r}}!\mathcal{N}_{2,\v{r}}!}{\left(\mathcal{N}_{1,\v{r}} + \mathcal{N}_{2,\v{r}} + 1\right)!}\right]
 \exp\Bigg(-\frac{g}{2}\sum_{\substack{c,c', \\ \mu, \v{r},\v{r}'}} I_{c,\mu,\v{r}} V_{\v{r},\v{r}'} I_{c',\mu,\v{r}'} \Bigg),
\end{equation}
\end{widetext}
which is a $J$-current formulation of the $\text{NCCP}^1$ model, see also Ref.~\onlinecite{Kuklov_PRL_2008}.

\bibliography{SU2refs}
\end{document}